\documentclass[12pt]{article}

\usepackage{graphics}
\usepackage{epsfig}
\usepackage{cite}
\input epsf

\title{Reconsideration of the inclusive prompt photon production at LHC with $k_T$-factorization}

\author{A.V.~Lipatov$^{1,\,2}$, M.A.~Malyshev$^1$}

\begin{document}

\maketitle

\begin{center}

{\it $^1$Skobeltsyn Institute of Nuclear Physics, Lomonosov Moscow State University, 119991 Moscow, Russia}\\
{\it $^2$Joint Institute for Nuclear Research, Dubna 141980, Moscow region, Russia}\\

\end{center}

\vspace{1cm}

\begin{center}

{\bf Abstract }

\end{center}

We reconsider the inclusive production of isolated 
prompt photons in $pp$ collisions at the LHC energies
in the framework of $k_T$-factorization approach.
Our analysis is based on the ${\cal O}(\alpha \alpha_s)$ off-shell (depending on the 
transverse momenta of initial quarks and gluons) 
production amplitudes of $q^* g^* \to \gamma q$ and $q^* \bar q^* \to \gamma g$
partonic subprocesses and transverse momentum dependent (or unintegrated) quark and gluon densities 
in a proton, which are chosen in accordance with the Kimber-Martin-Ryskin prescription.
We show that the sub-leading high-order ${\cal O}(\alpha \alpha_s^2)$ 
contributions, not covered by the non-collinear evolution of parton 
densities, are important to describe latest LHC data.

\vspace{2cm}

\noindent
PACS number(s): 12.38.-t, 13.85.-t

\newpage

Prompt photon production at hadron colliders is 
presently of considerable interest from both theoretical and experimental 
points of view\cite{1,2,3}.
It provides a direct probe of the hard subprocess dynamics because the 
produced photons are largely insensitive to the effects 
of final-state hadronization.
The measured cross sections are sensitive to the parton (quark and gluon) content 
of a proton since, at the leading order (LO), the prompt photons are produced mainly via quark-gluon 
Compton scattering or quark-antiquark annihilation.
Prompt photon production represents an important background to many processes 
involving photons in the final state, including Higgs boson production\cite{4}.
Therefore, it is essential to have accurate
QCD predictions for corresponding cross sections.

The CMS and ATLAS Collaborations have reported measurements\cite{1,2} of the
inclusive prompt photon production at the LHC energy $\sqrt s = 7$~TeV.
First measurements of inclusive photon cross sections at $\sqrt s = 8$~TeV
have been presented by the ATLAS Collaboration very recently\cite{3}.
These measurements extend the previous ones to wider ranges of 
photon pseudorapidity $\eta^\gamma$ and transverse energy $E_T^\gamma$, 
up to $E_T^\gamma \sim 1.5$~TeV. The pQCD predictions\cite{5,6} calculated 
at the next-to-leading order (NLO) agree with 
the LHC data within the theoretical and experimental uncertainties, although tend to 
underestimate the ATLAS data\cite{2} at $E_T^\gamma \sim 100$~GeV and 
overestimate the CMS data\cite{1} at lower $E_T^\gamma \sim 40$~GeV.
An alternative QCD description can be achieved in 
the framework of $k_T$-factorization approach\cite{7,8}, which is
based on the small-$x$ Balitsky-Fadin-Kuraev-Lipatov (BFKL)\cite{9} evolution equation
and provides solid theoretical grounds for the effects of initial gluon 
radiation and intrinsic parton transverse momentum. 
The latter
is known to be important for description of the prompt photon production
at hadron colliders\cite{10,11}, and the high-energy resummation formalism
was applied for photon production\cite{12,13}.

In the present note we give a systematic analysis of recent LHC data\cite{1,2,3}
using the $k_T$-factorization approach. 
Our consideration below is mainly based on the ${\cal O}(\alpha \alpha_s)$ off-shell (depending on the 
transverse momenta of initial quarks and gluons) 
quark-gluon Compton scatterring or quark-antiquark annihilation subprocesses.
We see certain advantages in the fact that, 
even with the LO partonic amplitudes, 
one can include a large piece of high-order corrections
(namely, part of NLO + NNLO terms and terms containing leading $\log 1/x$ enhancement of cross sections
due to real parton emissions in initial state, according to the BFKL evolution)
taking them into
account in the form of transverse momentum dependent (TMD) parton 
densities\footnote{A detailed description of the $k_T$-factorization
approach can be found, for example, in reviews\cite{14}.}.
It is known that such terms give the main contribution to the 
production cross section at high energies.
Unlike earlier calculations\cite{15,16,17,18,19}, to evaluate the 
off-shell production amplitudes we employ the reggeized parton approach\cite{20,21,22} 
based on the effective action formalism\cite{23},
that ensures the gauge invariance of
obtained amplitudes despite the off-shell initial quarks and
gluons\footnote{The investigation\cite{19} was based on the off-shell partonic 
amplitudes gauge-invariant in a small-$x$ limit.} and 
therefore significantly improves previous considerations\cite{15,16,17,18,19}.
We choose the TMD parton densities in a proton in accordance with the 
Kimber-Martin-Ryskin (KMR) prescription\cite{24}, currently explored at the NLO\cite{25},
and examine an assumption\cite{18} on the TMD sea quark 
densities in a proton applied in our previous consideration\cite{19}.
The numerical effect of this approximation is specially investigated below. 
In addition, 
we 
take into account some ${\cal O}(\alpha \alpha_s^2)$ contributions,
namely $q q^\prime \to \gamma q q^\prime$ 
ones.
The latter probe essential large $x$ (see below) and therefore
can be calculated in the traditional collinear QCD factorization scheme.
Thus, we rely on a combination of two techniques 
with each of them being used where it is most suitable.
The improvement of our previous predictions\cite{19} as it described above 
is a special goal of present note.

Let us start from a short review of calculation steps.
We describe first the evaluation
of the off-shell amplitudes of quark-gluon Compton scattering and 
quark-antiquark annihilation subprocesses:
\begin{equation}
  q^*(k_1) + g^*(k_2) \to \gamma(p_1) + q(p_2),
\end{equation}
\begin{equation}
  q^*(k_1) + \bar q^*(k_2) \to \gamma(p_1) + g(p_2),
\end{equation}

\noindent
where the four-momenta of corresponding particles are given in
the parentheses. 
In the center-of-mass frame of colliding protons, having
four-momenta $l_1$ and $l_2$, we define 
\begin{equation}
  k_1 = x_1 l_1 + k_{1T}, \quad k_2 = x_2 l_2 + k_{2T},
\end{equation}

\noindent
where $x_1$ and $x_2$ are the longitudinal 
momentum fractions of the protons 
carried by the interacting off-shell partons 
having transverse four-momenta $k_{1T}$ and $k_{2T}$ (note that
$k_{1T}^2 =  - {\mathbf k}_{1T}^2 \neq 0$, $k_{2T}^2 =  - {\mathbf k}_{2T}^2 \neq 0$).
In the reggeized parton approach the off-shell amplitude 
of subprocess~(1) reads:
\begin{equation}
  {\cal A}(q^* g^* \to \gamma q) = e e_q \, g \, t^a \epsilon^\mu(p_1) \epsilon^\nu(k_2) \, \bar v_{s_1}(p_2) \, {\cal A}^{\mu \nu}(q^* g^* \to \gamma q) \, u_{s_2}(x_1 l_1),
\end{equation}

\noindent
where $e$ and $e_q$ are the electron and quark (fractional) electric charges,
$g$ is the strong charge, $a$ is the eight-fold color index,
$\epsilon^\mu(p_1)$ and $\epsilon^\nu(k_2)$ are the 
polarization four-vectors and
\begin{equation}
  \displaystyle {\cal A}^{\mu \nu}(q^* g^* \to \gamma q) = \gamma^\nu {\hat k_1 - \hat p_1\over (k_1 - p_1)^2} \Gamma^\mu_{(+)}(k_1,p_1)
  \displaystyle + \gamma^\mu {\hat k_1 + \hat k_2\over (k_1 + k_2)^2} \Gamma^\nu_{(+)}(k_1,-k_2) \, + \atop
  \displaystyle { + \, \hat k_1 {l_1^\mu l_1^\nu \over (l_1 \cdot k_2) (l_1 \cdot p_1)}}.
\end{equation}

\noindent
The latter term in~(5) is the induced term, and 
we neglected the quark masses. The off-shell 
amplitude of subprocess~(2) reads:
\begin{equation}
  {\cal A}(q^* \bar q^* \to \gamma g) = e e_q \, g \, t^a \epsilon^\mu(p_1) \epsilon^\nu(p_2) \, \bar v_{s_1}(x_2 l_2) \, {\cal A}^{\mu \nu}(q^* \bar q^* \to \gamma g) \, u_{s_2}(x_1 l_1),
\end{equation}

\noindent
where
\begin{equation}
  \displaystyle {\cal A}^{\mu \nu}(q^* \bar q^* \to \gamma g) = \Gamma^\nu_{(-)}(k_2,p_2) {\hat k_1 - \hat p_1\over (k_1 - p_1)^2} \Gamma^\mu_{(+)}(k_1,p_1)
  \displaystyle + \Gamma^\mu_{(-)} {\hat k_1 - \hat p_2\over (k_1 - p_2)^2} \Gamma^\nu_{(+)}(k_1,k_2) \, + \atop
  \displaystyle { + \, \hat k_1 {l_1^\mu l_1^\nu \over (l_1 \cdot k_2) (l_1 \cdot p_1)} - \hat k_2 {l_2^\mu l_2^\nu \over (l_2 \cdot k_2) (l_2 \cdot p_1)}}.
\end{equation}

\noindent
The effective vertices read\cite{20,21}:
\begin{equation}
  \Gamma^\mu_{(+)}(k,q) = \gamma^\mu - \hat k {l_1^\mu \over (l_1 \cdot q)},
\end{equation}
\begin{equation}
  \Gamma^\mu_{(-)}(k,q) = \gamma^\mu - \hat k {l_2^\mu \over (l_2 \cdot q)},
\end{equation}

\noindent
The summation on the final state photon and gluon polarizations is carried out with the 
usual covariant formula:
\begin{equation}
  \sum \epsilon^\mu(p) \epsilon^{*\, \nu}(p) = - g^{\mu \nu}.
\end{equation}

\noindent
In contrast, according to the $k_T$-factorization prescription\cite{7,8}, 
the summation over the polarizations 
of incoming off-shell gluons is carried with
\begin{equation}
  \sum \epsilon^\mu(k) \epsilon^{*\, \nu}(k) = { {\mathbf k}_T^\mu {\mathbf k}_T^\nu \over {\mathbf k}_T^2}.
\end{equation}

\noindent
In the limit of collinear QCD factorization, when ${\mathbf k}_T^2 \to 0$, this expression converges to 
the ordinary one after averaging on the azimuthal angle.
The spin density matrix for all initial off-shell spinors 
in the parton reggezation approach is taken in the usual form:
\begin{equation}
  \sum u(x_i l_i) \bar u(x_i l_i) = x_i \hat l_i,
\end{equation}

\noindent
where $i = 1$ or $2$ and we omittеd the spinor indices. Further calculations are 
straightforward and in other respects follow the standard QCD Feynman 
rules. The evaluation of traces was performed using the algebraic 
manipulation system \textsc{form}\cite{26}.

To calculate the contributions of subprocesses~(1) and (2) to the prompt photon production 
cross section we have to convolute the relevant partonic cross sections and the 
TMD parton densities in a proton:
\begin{equation}
  \displaystyle \sigma(pp \to \gamma + X) = \sum_{a,b} \int {1 \over 16 \pi (x_1 x_2 s)^2 } f_a(x_1,{\mathbf k}_{1T}^2,\mu_F^2) f_b(x_2,{\mathbf k}_{2T}^2,\mu_F^2) \times \atop { 
  \displaystyle  \times \, |\bar {\cal A} (a^* b^* \to \gamma c)|^2 d{\mathbf k}_{1T}^2 d{\mathbf k}_{2T}^2 d{\mathbf p}_{1T}^2 dy_1 dy_2 {d\phi_1 \over 2\pi} {d\phi_2 \over 2\pi}},
\end{equation}

\noindent
where $a$, $b$ and $c$ are parton indices ($q$ or $g$), $f_a(x,{\mathbf k}_{T}^2,\mu_F^2)$ are
the TMD parton densities at the factorization scale $\mu_F$, $s$ is the total energy, $y_1$ and $y_2$
are the center-of-mass rapidities of final state particles, and $\phi_1$ and $\phi_2$ are the
azimuthal angles of initial partons.

As it was mentioned above,
we take into account additional contribution from 
$q q^\prime \to \gamma q q^\prime$
subprocess. We apply here the collinear limit of formulas
obtained earlier\cite{18,19}.

It is well-known that the photon production cross section suffers from a 
final state divergence when the photon becomes collinear to the
outgoing parton. This collinear divergence cannot be removed by
adding the virtual corrections and is usually absorbed into the 
parton-to-photon fragmentation functions.
In the present note we used an approach proposed in\cite{18}.
So, the standard QCD perturbation theory can be only applied when the wavelength of the emitted 
photon (in the emitting quark rest frame) becomes larger than the 
typical hadronic scale ${\cal O}(1$~GeV$^{-1})$. Below this scale,
the non-perturbative effects of photon fragmentation 
take place
and have to be taken into account.
Accordingly, we split the photon cross section into two pieces:
\begin{equation}
  \sigma = \sigma_{\rm pert}(\mu_{\rm reg}^2) + \sigma_{\rm non-pert}(\mu_{\rm reg}^2),
\end{equation}

\noindent
where $\sigma_{\rm pert}(\mu_{\rm reg}^2)$ is the perturbative contribution
and $\sigma_{\rm non-pert}(\mu_{\rm reg}^2)$ is the non-perturbative one which 
includes the fragmentation component. Both of them depend on the regularization 
scale $\mu_{\rm reg}$, 
which can be used to separate these two pieces.
Following\cite{18}, 
we restrict $\sigma_{\rm pert}(\mu_{\rm reg}^2)$ to 
the region $M \geq \mu_{\rm reg}$,
where $M$ is the invariant mass of the photon + parton subsystem
and $\mu_{\rm reg} \sim 1$~GeV is the typical hadronic scale. 
Under this condition, the contribution
$\sigma_{\rm pert}(\mu_{\rm reg}^2)$ is free from collinear divergences.
The sensitivity of our results to the choice of $\mu_{\rm reg}$ is reasonably soft\footnote{Under the isolation condition, see below.}
and investigated below. 

Next, the size of conventional fragmentation contribution
is dramatically reduced by the photon
isolation criterion introduced in the experimental analyses\cite{1,2,3},
mainly to reduce huge background.
This criterion is the following: a photon is isolated if the amount 
of hadronic transverse energy $E_T^{\rm had}$ deposited inside a cone with 
aperture $R$ centered around the photon direction in the 
pseudo-rapidity and azimuthal angle plane, is smaller than
some value $E_T^{\rm max}$:
\begin{equation}
  \displaystyle E_T^{\rm had} \leq E_T^{\rm max}, \atop {
  \displaystyle (\eta^{\rm had} - \eta^\gamma)^2 + (\phi^{\rm had} - \phi^\gamma)^2 \leq R^2 }.
\end{equation}

\noindent
The CMS Collaboration takes $R = 0.4$ and $E_T^{\rm max} = 5$~GeV\cite{1}, whereas
the ATLAS Collaboration applies 
$E_T^{\rm max} = 7$~GeV\cite{2} or $E_T^{\rm max} = 4.8$~GeV + $4.2 \cdot 10^{-3} \times E_T^\gamma$\cite{3} with the same $R$.
According to the estimates\cite{1,2,3},
after applying the isolation cut the fragmentation contribution 
amounts to about 10\% of the measured cross section.
This value is smaller than the typical theoretical uncertainties in calculating 
the perturbative contribution $\sigma_{\rm pert}(\mu_{\rm reg}^2)$. 
Moreover, the isolation criterion~(15), applied in our calculations, 
is used as a tool to remove the 
non-perturbative part of cross section~(14), where final photon
is radiated close to quark (inside the isolation cone).


To calculate the TMD parton densities in a proton we adopt the  
KMR prescription\cite{24} developed at the NLO\cite{25}.
The KMR approach is a formalism to construct the TMD 
parton densities from the known conventional parton distributions. 
The key assumption is that the
$k_T$ dependence enters 
at the last evolution step, so that the 
Dokshitzer-Gribov-Lipatov-Altarelli-Parisi (DGLAP) evolution\cite{27} 
can be used up to this step. 
Numerically, for the input we applied parton densities
from the MSTW'2008 NLO set\cite{28}.

Other essential parameters we take as follows: renormalization 
and factorization scales $\mu_R = \mu_F = \xi E_T^\gamma$, where
the unphysical parameter $\xi$ is varied between $1/2$ and $2$ 
about the default value $\xi = 1$ to estimate the scale
uncertainties of our calculations.
The uncertainties originating from the 
cut-off parameter $M$
are estimated in the same way, by varying $M$ between
$0.5 < M < 2$~GeV about the default value $M = 1$~GeV.
We apply the two-loop formula for the strong coupling constant
with $n_f = 5$ active quark flavours at $\Lambda_{\rm QCD} = 226.2$~MeV
and use the running QED coupling constant over a 
wide region of $E_T^\gamma$, as it is measured by the ATLAS Collaboration.
The same renormalization scale $\mu_R$ is applied for both the QCD and QED coupling constants.
Everywhere the multidimensional integration have been performed
by the means of Monte Carlo technique, 
using the routine \textsc{vegas}\cite{29}.

We now are in a position to present our numerical results
in comparison with the LHC data\cite{1,2,3}.
So, the CMS Collaboration
has measured the prompt photon production cross section as a 
function of the photon transverse energy $E_T^\gamma$
in the kinematical 
region defined by $25 < E_T^\gamma < 400$~GeV and $|\eta^\gamma| < 2.5$ at $\sqrt s = 7$~TeV\cite{1}.
The ATLAS Collaboration has measured 
the photon cross sections as a functions
of transverse energy and pseudorapidity in the kinematic 
range $100 < E_T^\gamma < 1000$~GeV, $|\eta^\gamma| < 1.37$ and 
$1.52 < |\eta^\gamma| < 2.37$ at $\sqrt s = 7$~TeV\cite{2}.
Recently, the data taken at $\sqrt s = 8$~TeV
in the kinematic range $25 < E_T^\gamma < 1500$~GeV, $|\eta^\gamma| < 0.6$,
$0.6 < |\eta^\gamma| < 1.37$, $1.56 < |\eta^\gamma| < 1.81$ and 
$1.81 < |\eta^\gamma| < 2.37$ were presented by the ATLAS Collaboration\cite{3}.
The results of our calculations are shown in Figs. 1 --- 7.
In Figs.~1,~2 and~4 we confront the cross sections calculated 
as a function of $E_T^\gamma$
with the LHC data and plot corresponding 
data/theory ratios.
As one can see from Fig.~2, our results  
agree well with the ATLAS data taken at $\sqrt s = 7$~TeV and central pseudorapidities $|\eta^\gamma| < 1.37$
in the whole $E_T^\gamma$ region
within the experimental and theoretical
uncertainties. 
At $\sqrt s = 8$~TeV, perfect agreement with the recent ATLAS data for all $E_T^\gamma$
is achieved at $|\eta^\gamma| < 0.6$, see Fig.~4. In the next pseudorapidity subdivision, 
$0.6 < |\eta^\gamma| < 1.37$, the overall 
description of the data is rather satisfactory, although a some tendence to slightly underestimate the data 
at high $E_T^\gamma > 200$~GeV can be seen.
In the forward region, where $1.52 < |\eta^\gamma| < 2.37$, our predictions
lie somewhat below the ATLAS data, for both $\sqrt s = 7$ and 8~TeV.
This becomes clearer in the $\eta^\gamma$
distributions, presented by the ATLAS Collaboration for the first time
(see Fig.~3). The observed discrepancy 
could be attributed to the missing higher-order
contributions, not taken into account in our consideration.
However, we note that the ATLAS data 
in these subdivisions of $\eta^\gamma$
are close to the upper bound of 
theoretical uncertainties.
The CMS data 
are more or less well described for all pseudorapidities $\eta^\gamma$ (see Fig.~1),
although our predictions tend to slightly overestimate the
data at low $E_T^\gamma$ and underestimate them at high $E_T^\gamma$,
that could be due to the TMD parton
densities, involved in the calculations.
In the forward kinematical region, where $2.1 < \eta^\gamma < 2.5$,
the CMS data are described better compared to the ATLAS ones.


Let us turn to comparison of obtained results  
with the predictions based on a special assumption\cite{18} on the 
TMD sea quark density in a proton, which was used in our 
previous consideration\cite{19}.
The proposed scheme is based on the separation of
the TMD sea quark densities to the 
sea quarks appearing at the last step of
the gluon evolution and ones coming from the earlier (second-to-last,
third-to-last and other) gluon splittings.
First of them are calculated using ${\cal O}(\alpha \alpha_s^2)$
off-shell gluon-gluon fusion subprocess, $g^*g^* \to \gamma q \bar q$.
To estimate the second contributions the specific
properties of the KMR formalism, which enables us to discriminate 
between the various components of the TMD quark densities (see\cite{18,19}), are used.
The predictions based on this scheme are shown in Figs.~1 --- 4 by the dashed curves. 
We find that these predictions reproduce well
the recent LHC data\cite{1,2,3} at the central rapidities
(that agrees with the conclusions given in\cite{19})
and underestimate them in a forward region.
They lie somewhat below the newly presented calculations,
although both of them are rather close to each other and, in general, 
coincide within the theoretical uncertainties. 
Nevertheless, the latter describe better the latest ATLAS data\cite{2,3}.

The relative contributions to prompt photon production cross sections are 
shown in Figs.~5 --- 7. As it was expected, the off-shell quark-gluon 
Compton scattering subprocess dominates at low and moderate 
photon transverse energies.
The ${\cal O}(\alpha \alpha_s^2)$ contributions
from $q q^\prime \to \gamma q q^\prime$
subprocess
play a role mainly at high $E_T^\gamma$, where the 
large-$x$ region is probed. It supports our assumptions
that these subprocesses can be safely taken into account 
in the framework of collinear QCD factorization, thus avoiding an 
unnecessary complications of consideration.
However, these terms are important to describe the data.
The contribution of off-shell quark-antiquark annihilation
is negligible at the LHC conditions.

To conclude, we presented here analysis of latest 
LHC data on the inclusive prompt photon production at $\sqrt s = 7$ and $8$~TeV
in the framework of $k_T$-factorization approach.
Unlike previous studies, our consideration was
based on the ${\cal O}(\alpha \alpha_s)$ off-shell partonic amplitudes calculated
in the reggeized parton approach, that ensures their exact gauge 
invariance even with the off-shell initial partons.
In this way, even with the LO hard scaterring amplitudes, 
we include a large piece of high-order QCD corrections
taking them into account in the form of TMD parton densities.
To be precise, in the framework of KMR prescription used, we include
the NLO terms containing $\log 1/x$ enhancement of the 
cross section connected with the initial-state real parton emissions.
Such terms are known as giving the main high-order corrections to the cross section
at high energies\footnote{The
part of collinear NNLO pQCD corrections, namely, $\log 1/x$-enhanced terms, are effectively taken into account
in the calculations based on the scheme\cite{18,19}.}.
Of course, other high-order contributions, like virtual radiative corrections,
are not taken into account in our approach.
We achieved reasonably good agreement between our predictions 
and the CMS data for $E_T^\gamma \leq 100$~GeV in the whole region of photon  
pseudorapidity, $|\eta^\gamma| < 2.5$. 
At higher $E_T^\gamma$, our predictions tend to 
underestimate the CMS data.
The ATLAS data are described well
in the central pseudorapidity region, where $|\eta^\gamma| < 1.37$.
We showed that the sub-leading higher-order ${\cal O}(\alpha \alpha_s^2)$ 
contributions, not covered by the non-collinear parton evolution, 
are important to describe the LHC data, especially at high $E_T^\gamma$.
We examined the numerical effect of the special
assumption\cite{18} on the TMD sea quark 
densities in a proton used in the previous consideration\cite{19},
and found that our newly presented results describe a little better the 
latest ATLAS data\cite{2,3}.

{\sl Acknowledgements.} 
The authors are very grateful to S.~Baranov and H.~Jung for very useful discussions
and important remarks. This work was supported in part by RFBR grant 16-32-00176-mol-a,
grant of the President of Russian Federation NS-7989.2016.2 and
by the DESY Directorate in the framework of Moscow-DESY project on Monte-Carlo 
implementations for HERA-LHC.

\newpage

\begin{figure}
\begin{center}
\epsfig{figure=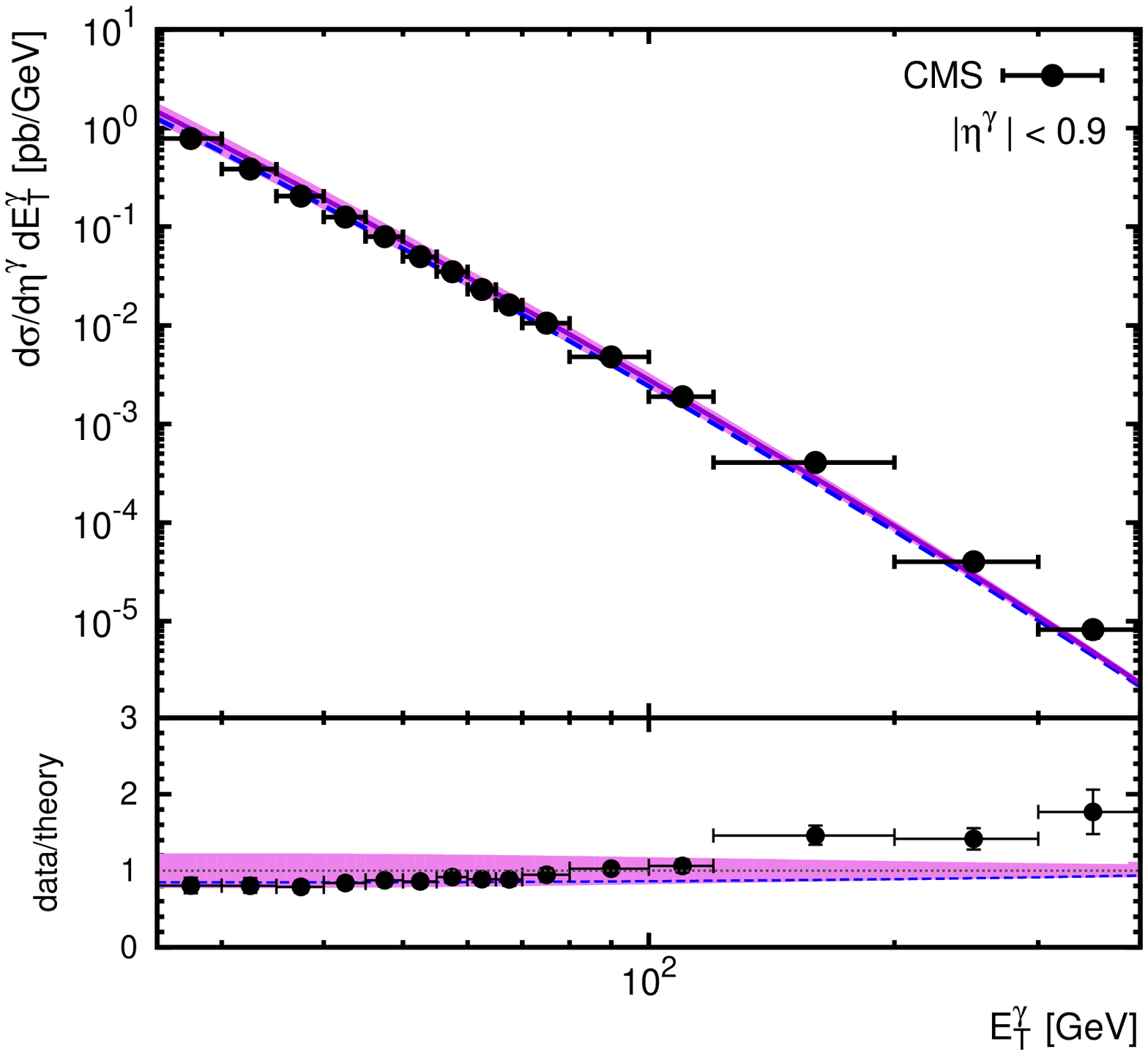, width = 6cm, angle = 270} 
\vspace{0.7cm} \hspace{-1cm}
\epsfig{figure=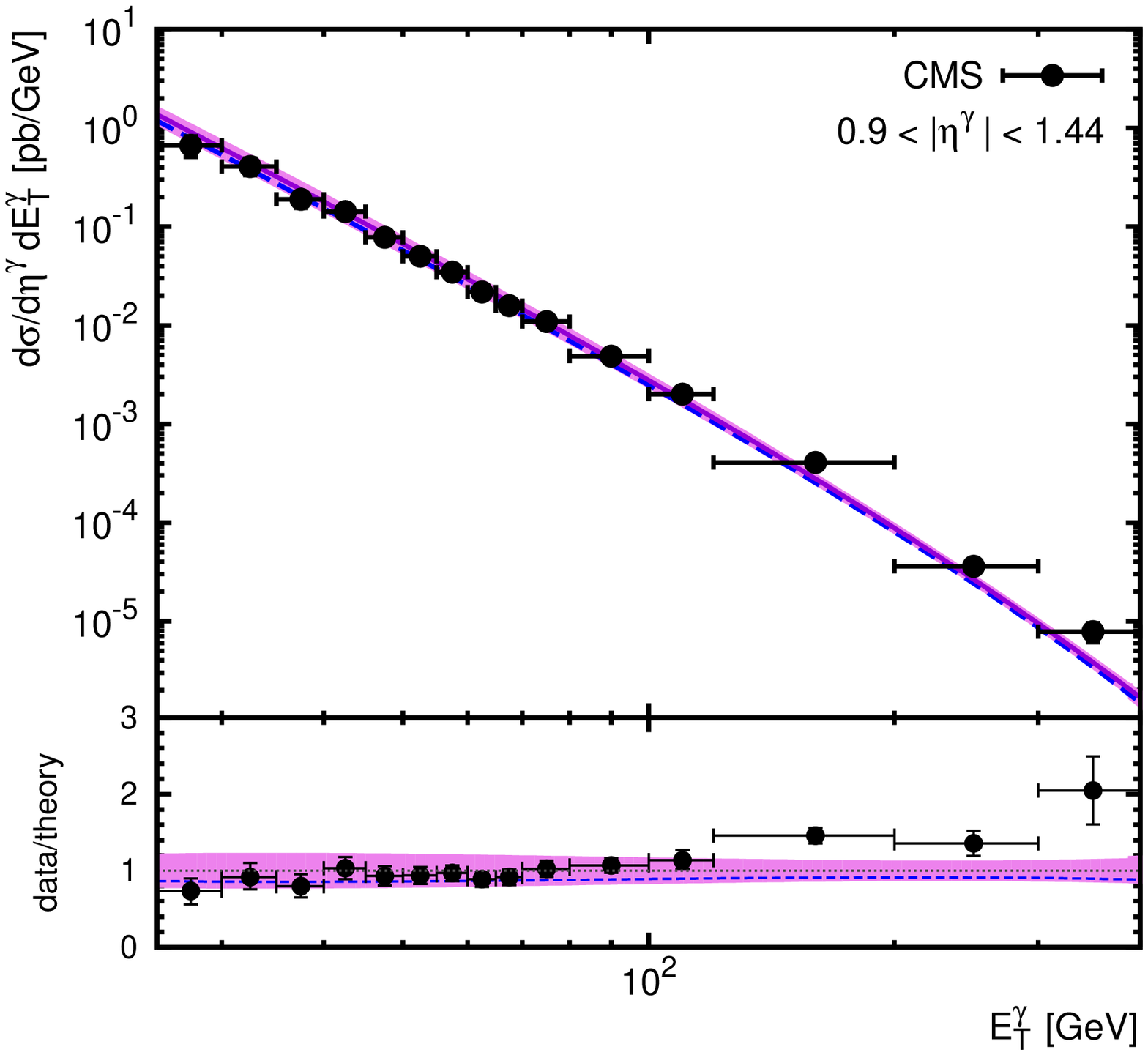, width = 6cm, angle = 270} 
\vspace{0.7cm}
\epsfig{figure=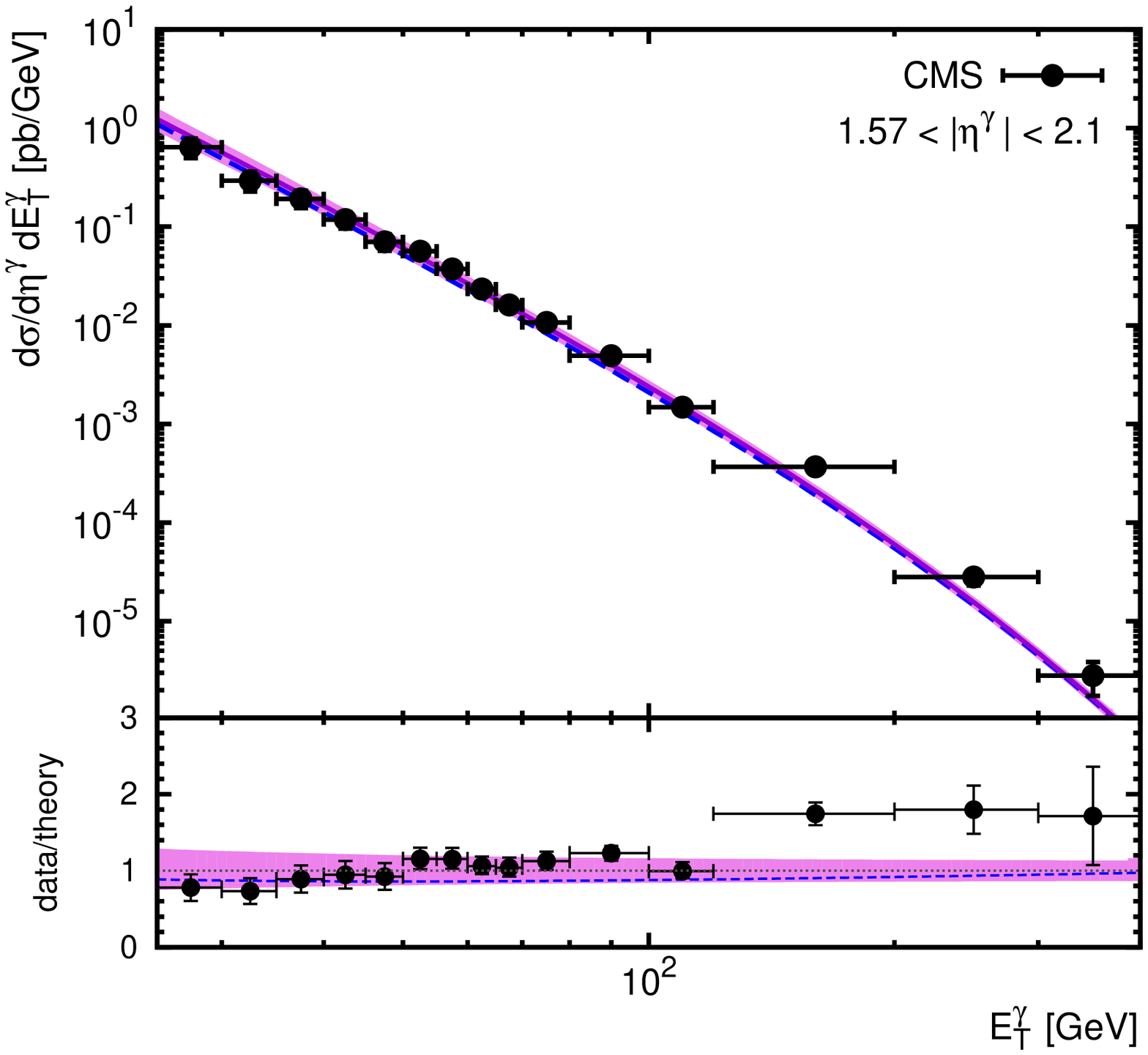, width = 6cm, angle = 270}
\hspace{-1cm}
\epsfig{figure=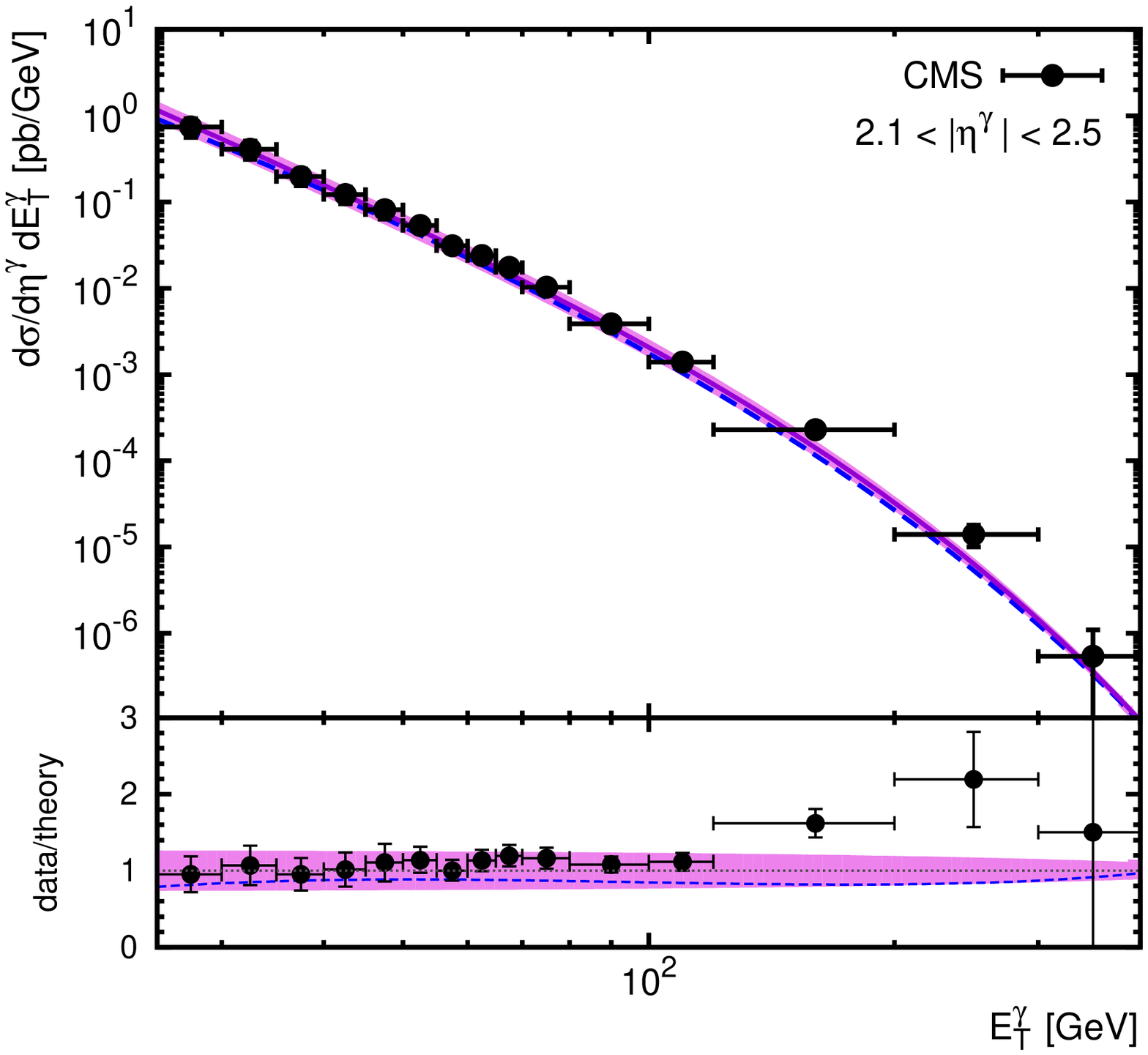, width = 6cm, angle = 270}
\caption{The inclusive prompt photon production at the LHC calculated
as a function of photon transverse energy $E_T^\gamma$ at $\sqrt s = 7$~TeV.
The solid curves correspond to the predictions obtained with the KMR parton 
densities at the default scale. The shaded band corresponds to 
the variation in scales $\mu_R$, $\mu_F$ and in parameter $\mu_{\rm reg}$, 
as described in the text. The dashed curves correspond to the special
assumption\cite{18} on the TMD sea quark density, applied as it was done in\cite{19}.
The experimental data are from CMS\cite{1}.}
\label{fig1}
\end{center}
\end{figure}

\begin{figure}
\begin{center}
\epsfig{figure=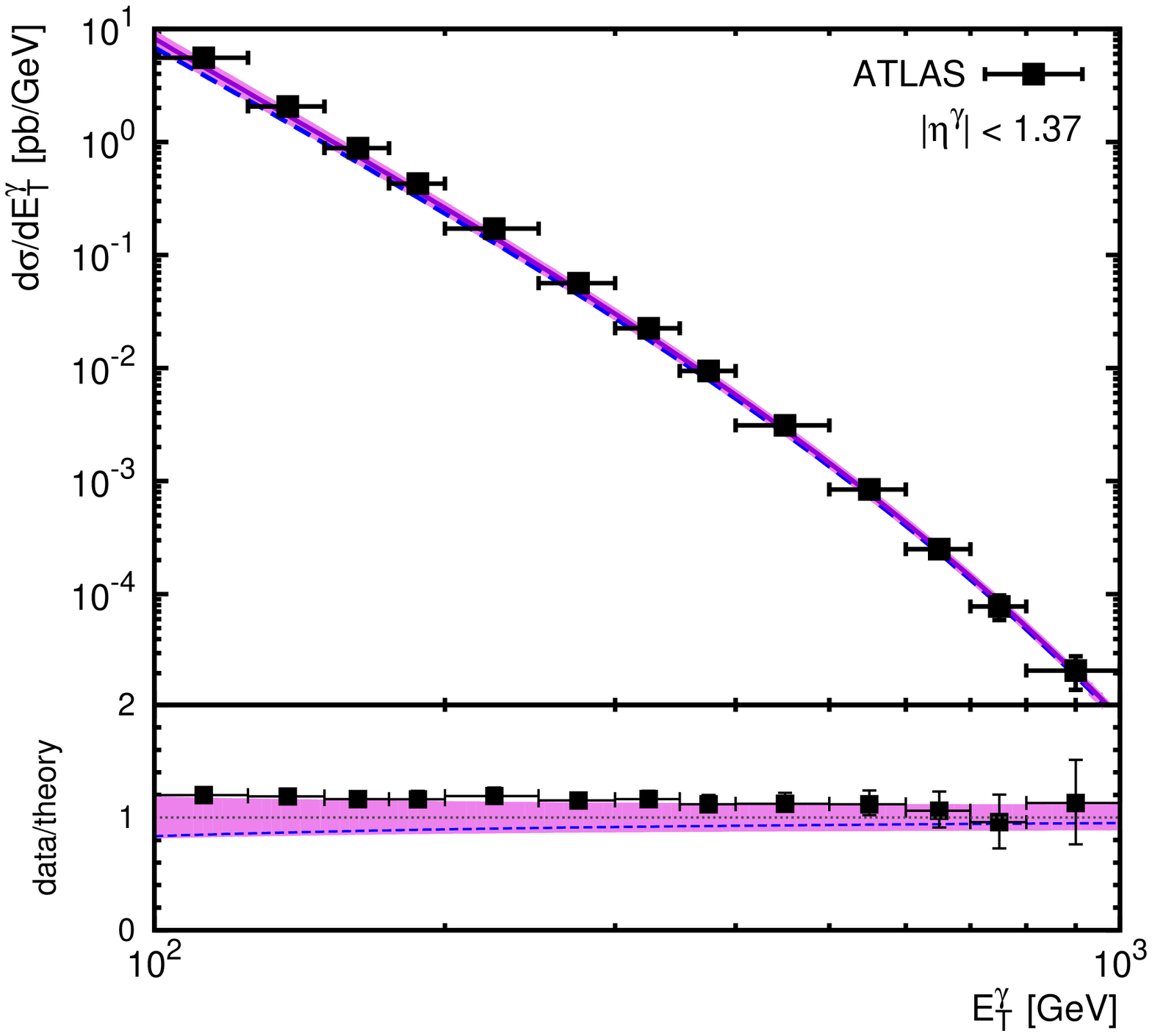, width = 6cm, angle = 270}
\hspace{-1cm}
\epsfig{figure=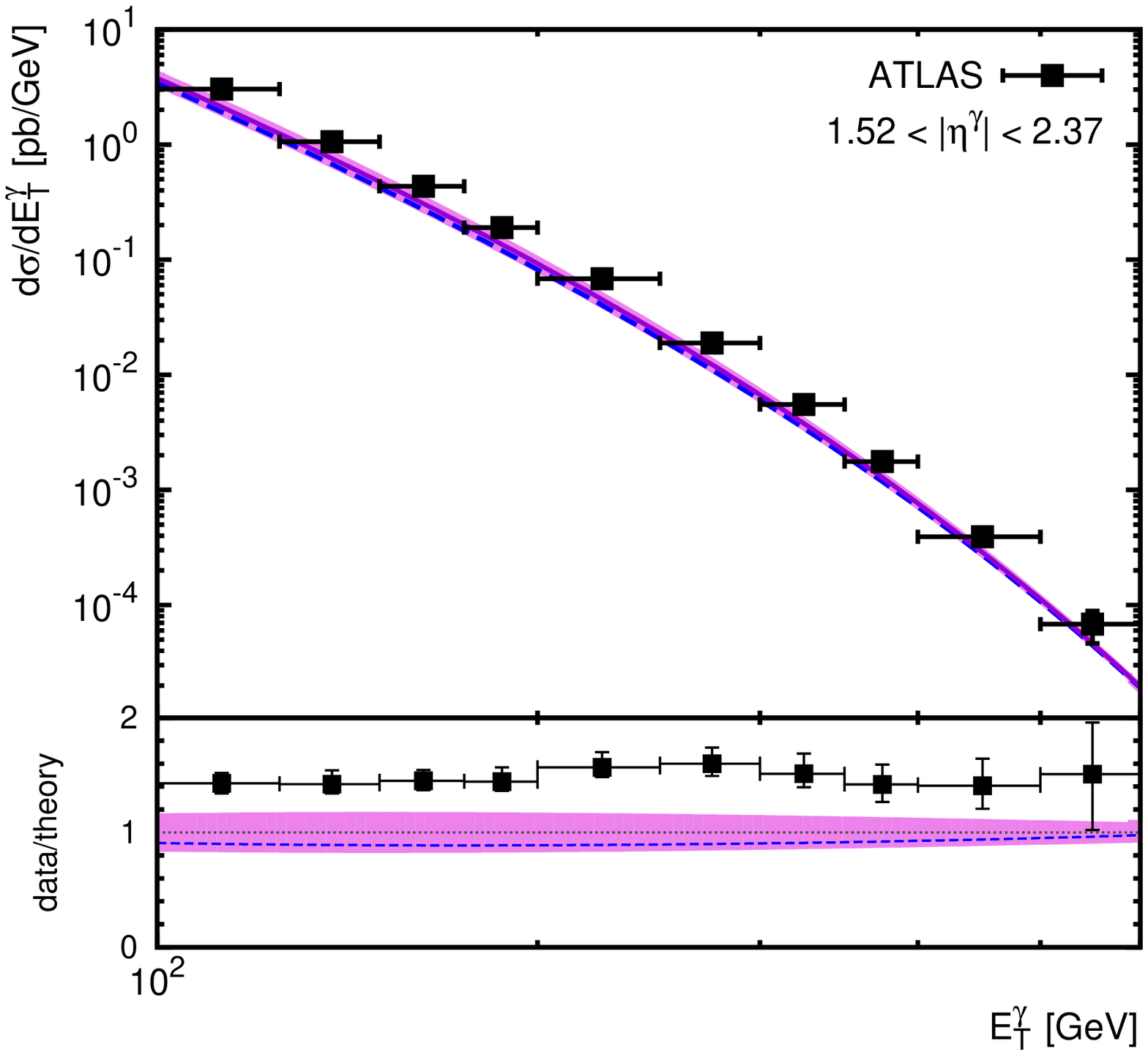, width = 6cm, angle = 270}
\caption{The inclusive prompt photon production at the LHC calculated
as a function of photon transverse energy $E_T^\gamma$ at $\sqrt s = 7$~TeV. 
Notation of all curves is the same as in Fig.~1. 
The experimental data are from ATLAS\cite{2}.}
\label{fig2}
\end{center}
\end{figure}

\begin{figure}
\epsfig{figure=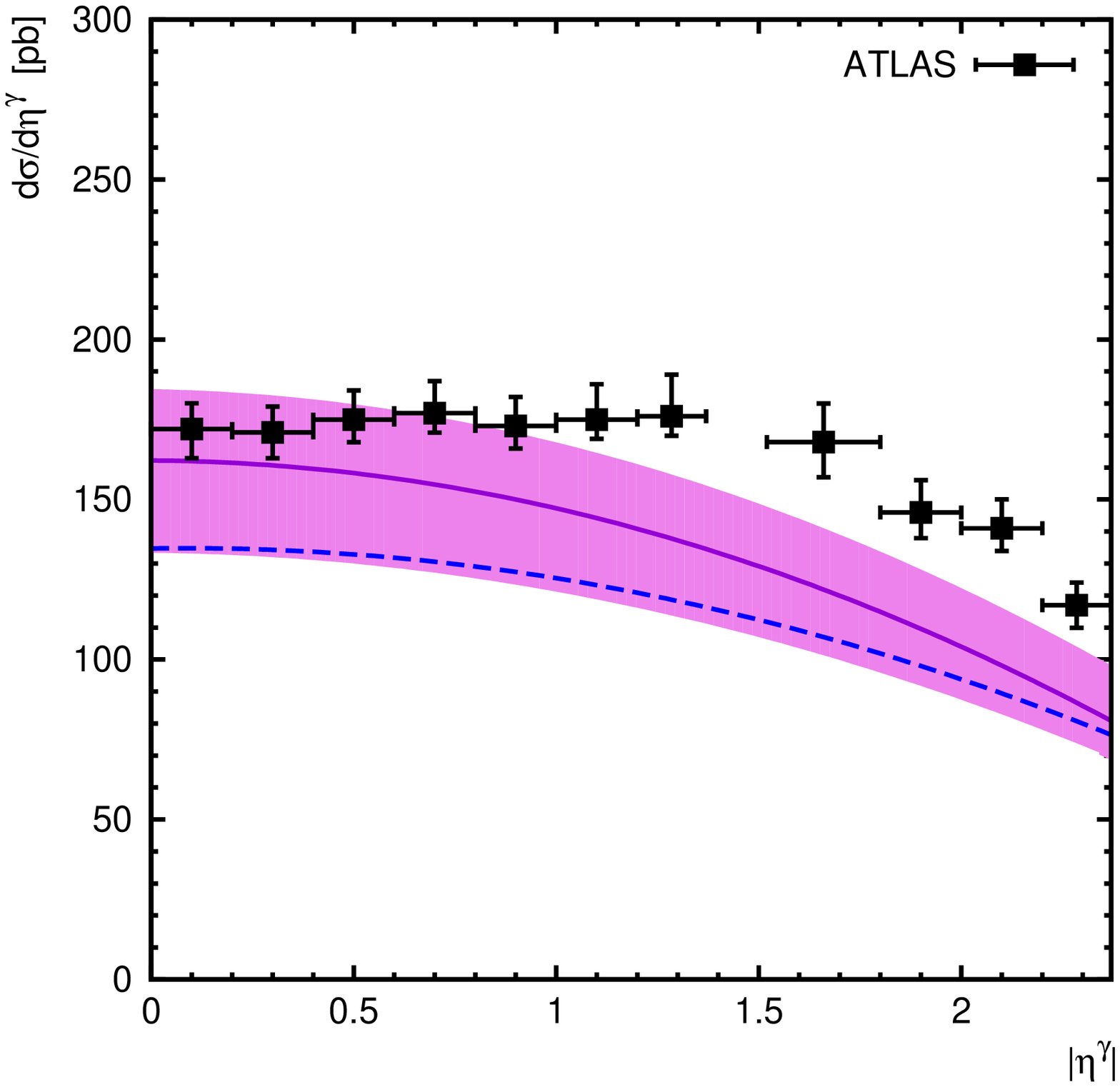, width = 5.8cm, angle = 270}
\hspace{-1cm}
\epsfig{figure=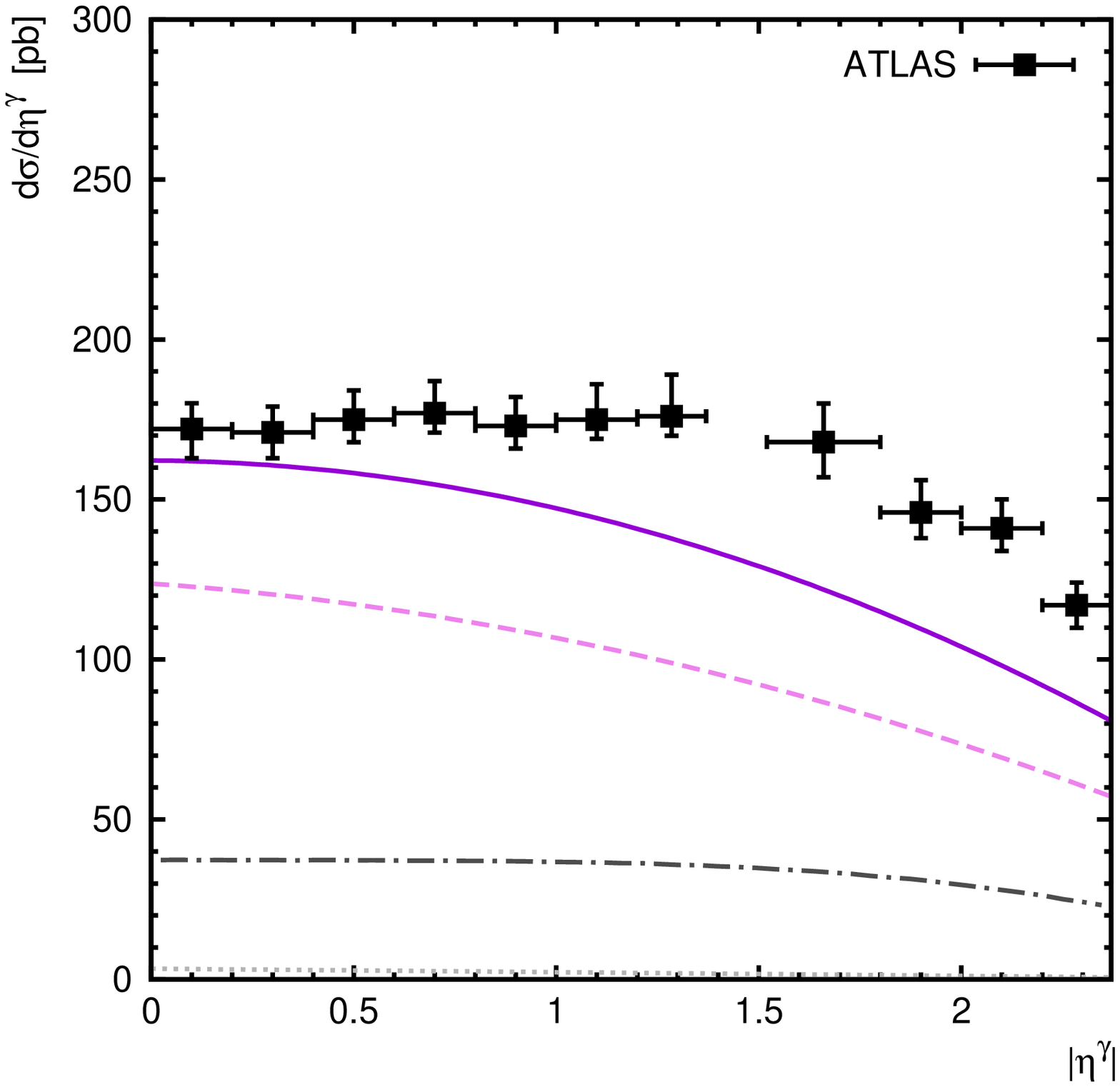, width = 5.8cm, angle = 270}
\caption{The inclusive prompt photon production at the LHC calculated
as a function of photon pseudorapidity $\eta^\gamma$ at $\sqrt s = 7$~TeV. 
Different contributions are shown on the right panel.
Notation of all curves in the left panel is the same as in Fig.~1.
The dashed, dotted and dash-dotted 
curves in the right panel correspond to the $q^* g^* \to \gamma q$, $q^* \bar q^* \to \gamma g$ and
$q q^\prime \to \gamma q q^\prime$ subprocesses, while
the solid curves represent the sum of all contributions.
The experimental data are from ATLAS\cite{2}.}
\label{fig3}
\end{figure}

\begin{figure}
\begin{center}
\epsfig{figure=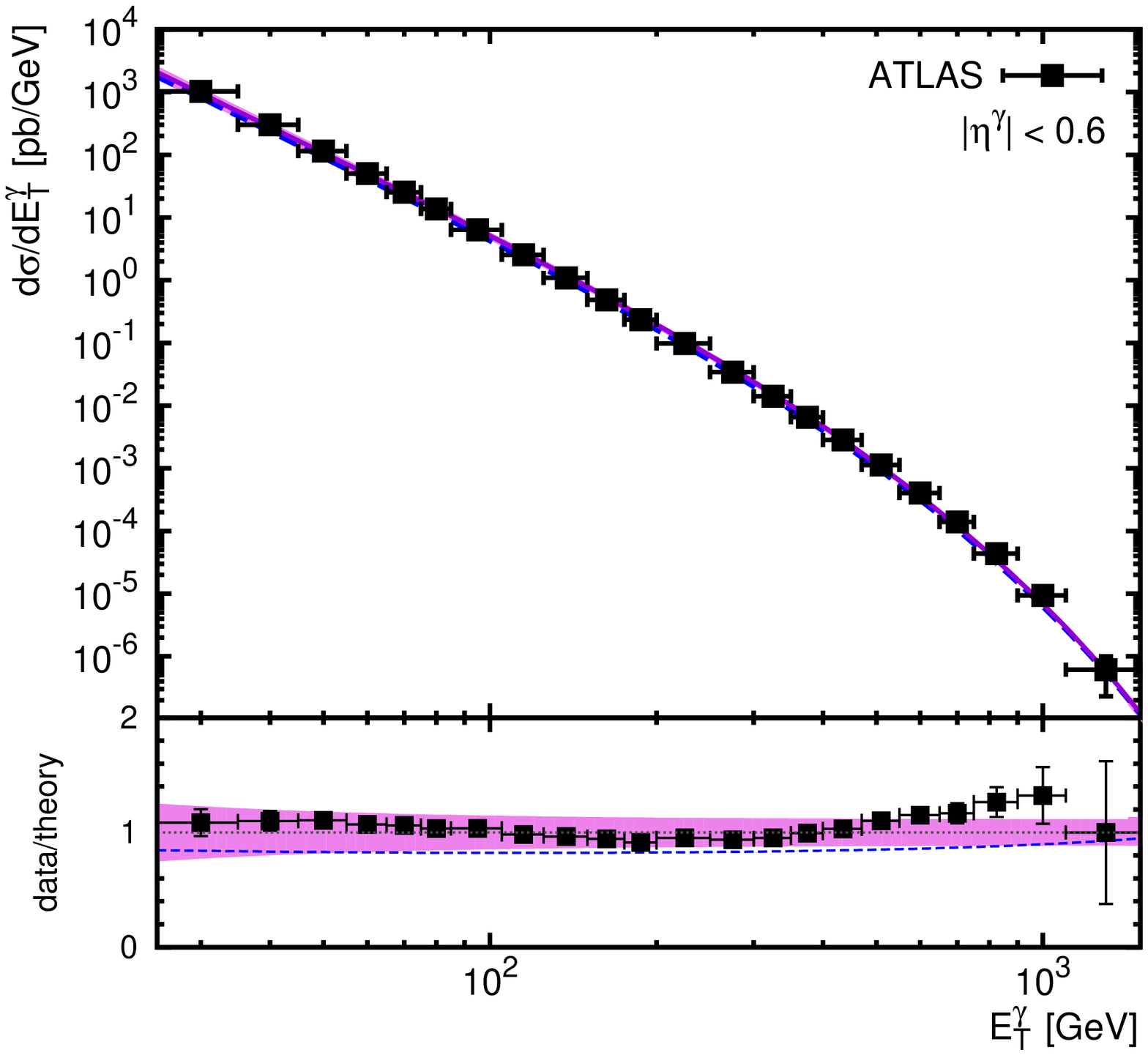, width = 6cm, angle = 270} 
\vspace{0.7cm} \hspace{-1cm}
\epsfig{figure=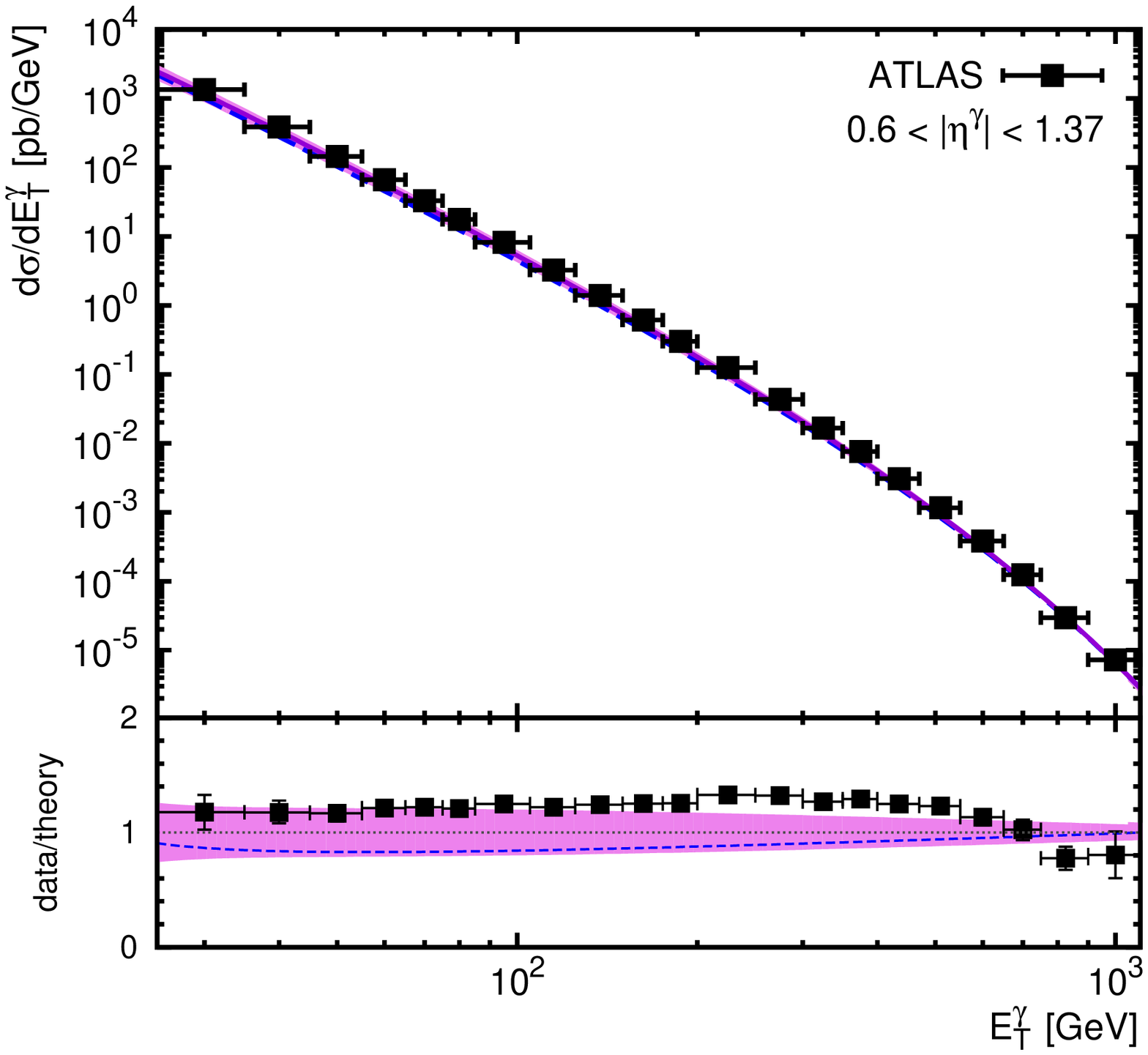, width = 6cm, angle = 270} 
\vspace{0.7cm}
\epsfig{figure=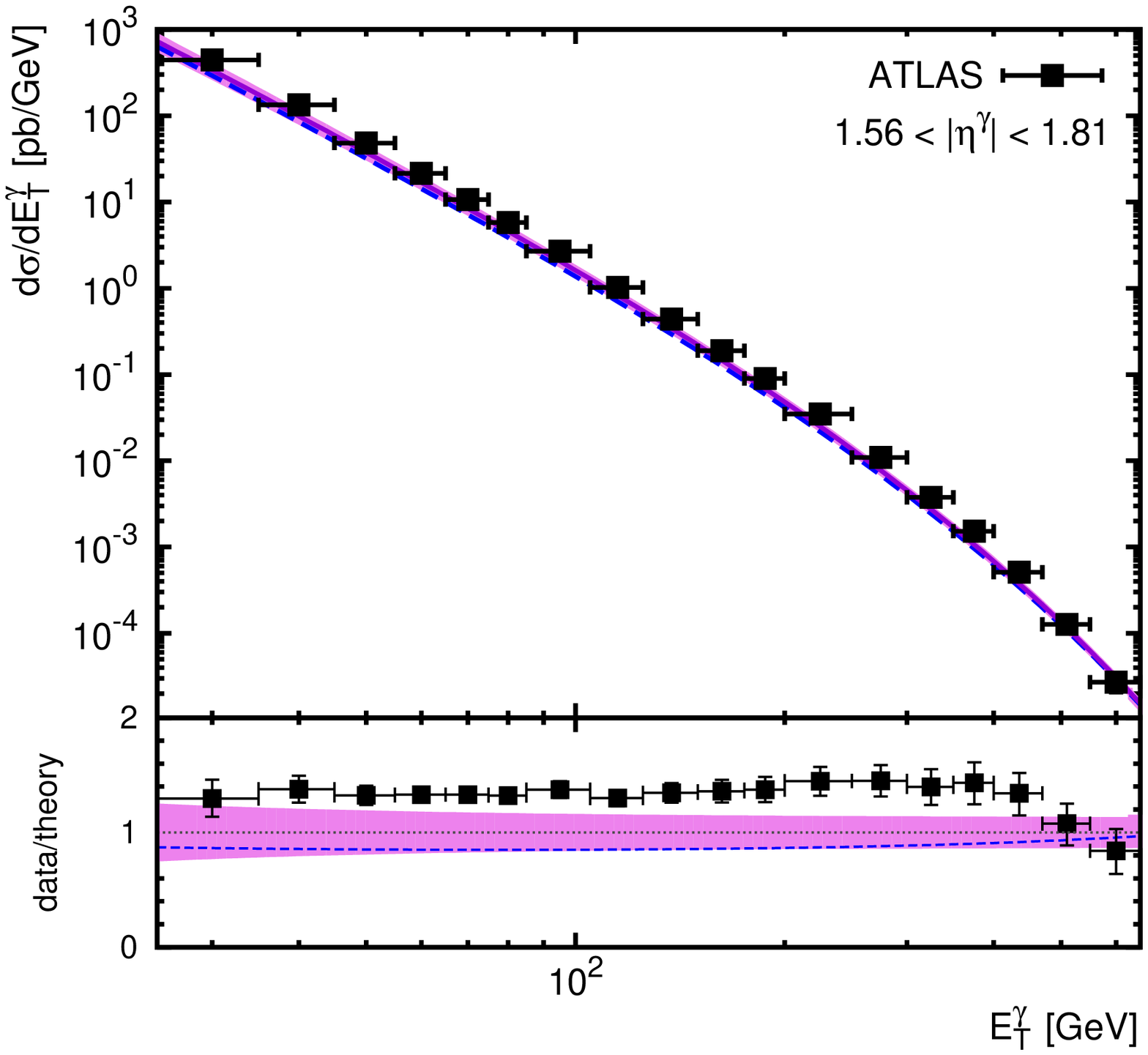, width = 6cm, angle = 270}
\hspace{-1cm}
\epsfig{figure=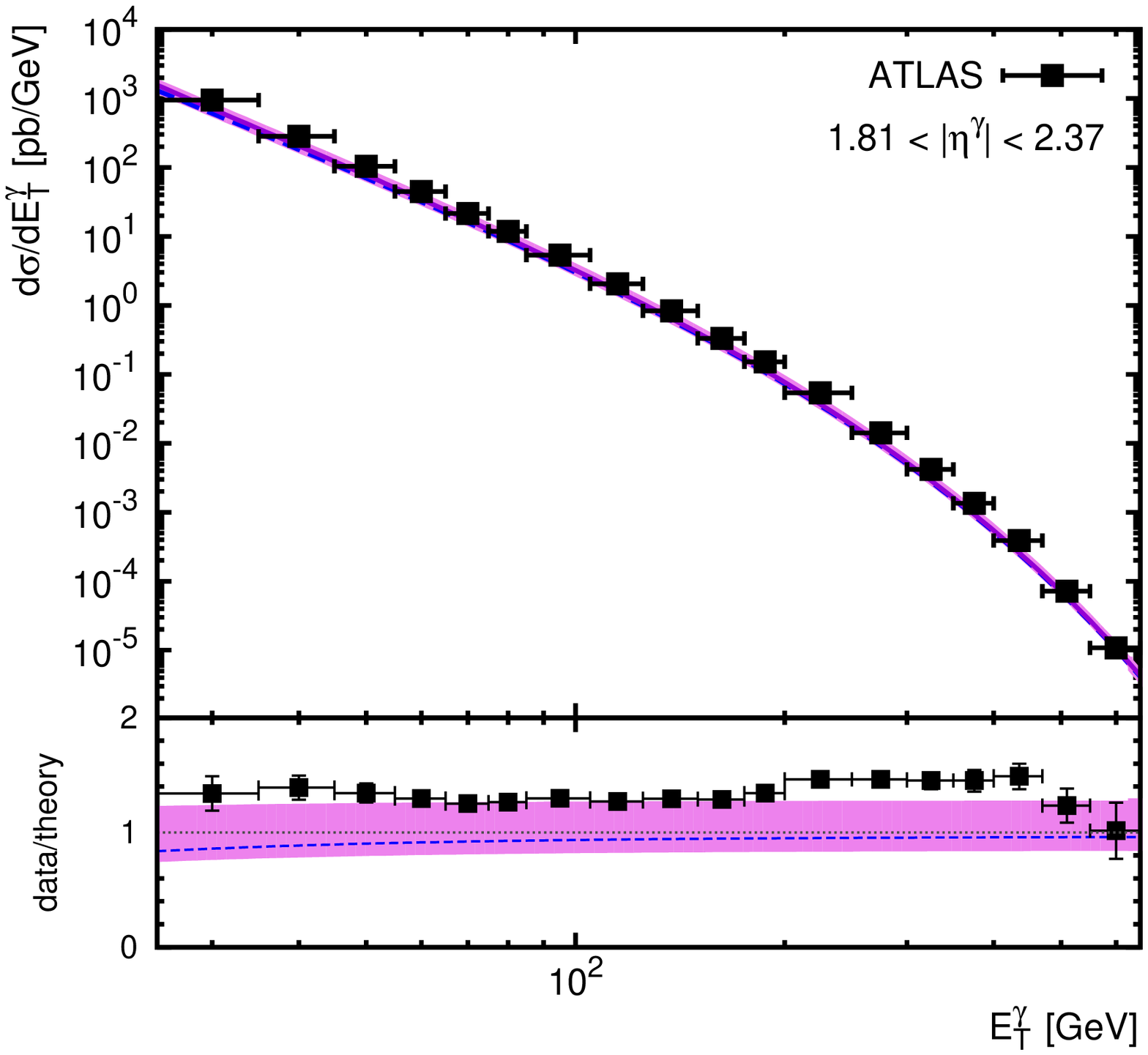, width = 6cm, angle = 270}
\caption{The inclusive prompt photon production at the LHC calculated
as a function of photon transverse energy $E_T^\gamma$ at $\sqrt s = 8$~TeV. 
Notation of all curves is the same as in Fig.~1. 
The experimental data are from ATLAS\cite{3}.}
\label{fig4}
\end{center}
\end{figure}

\begin{figure}
\begin{center}
\epsfig{figure=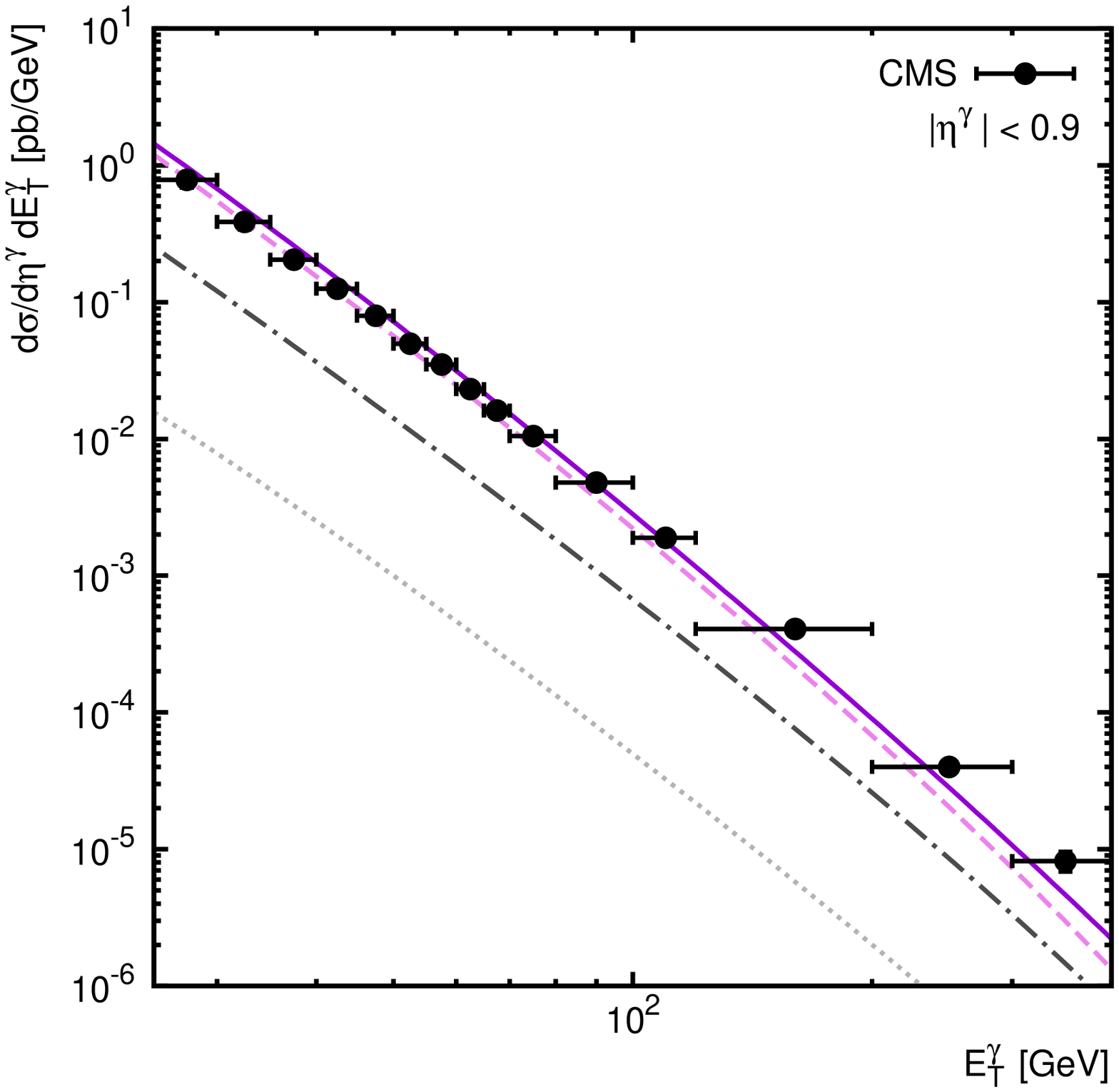, width = 6cm, angle = 270} 
\vspace{0.7cm} \hspace{-1cm}
\epsfig{figure=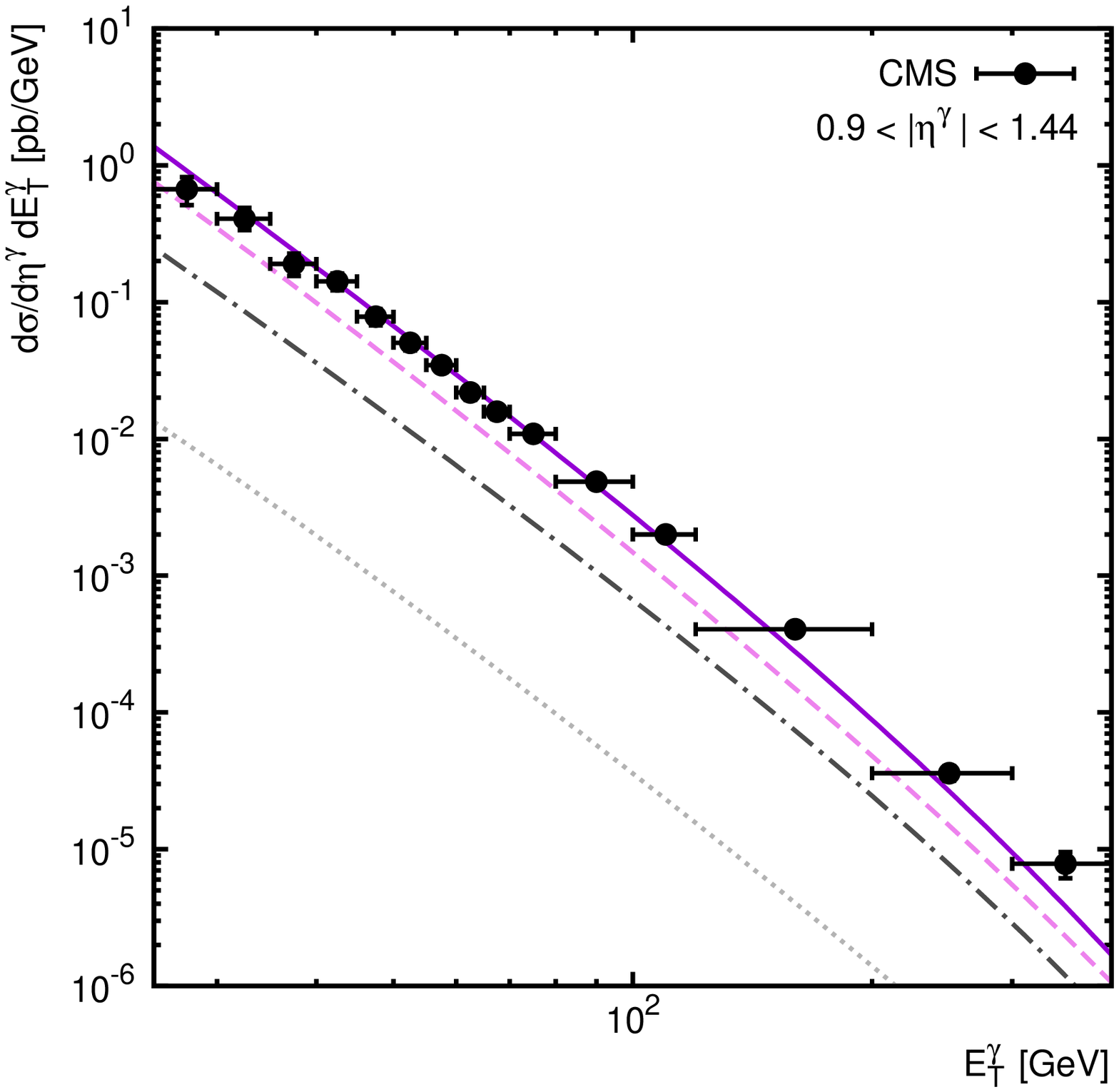, width = 6cm, angle = 270} 
\vspace{0.7cm}
\epsfig{figure=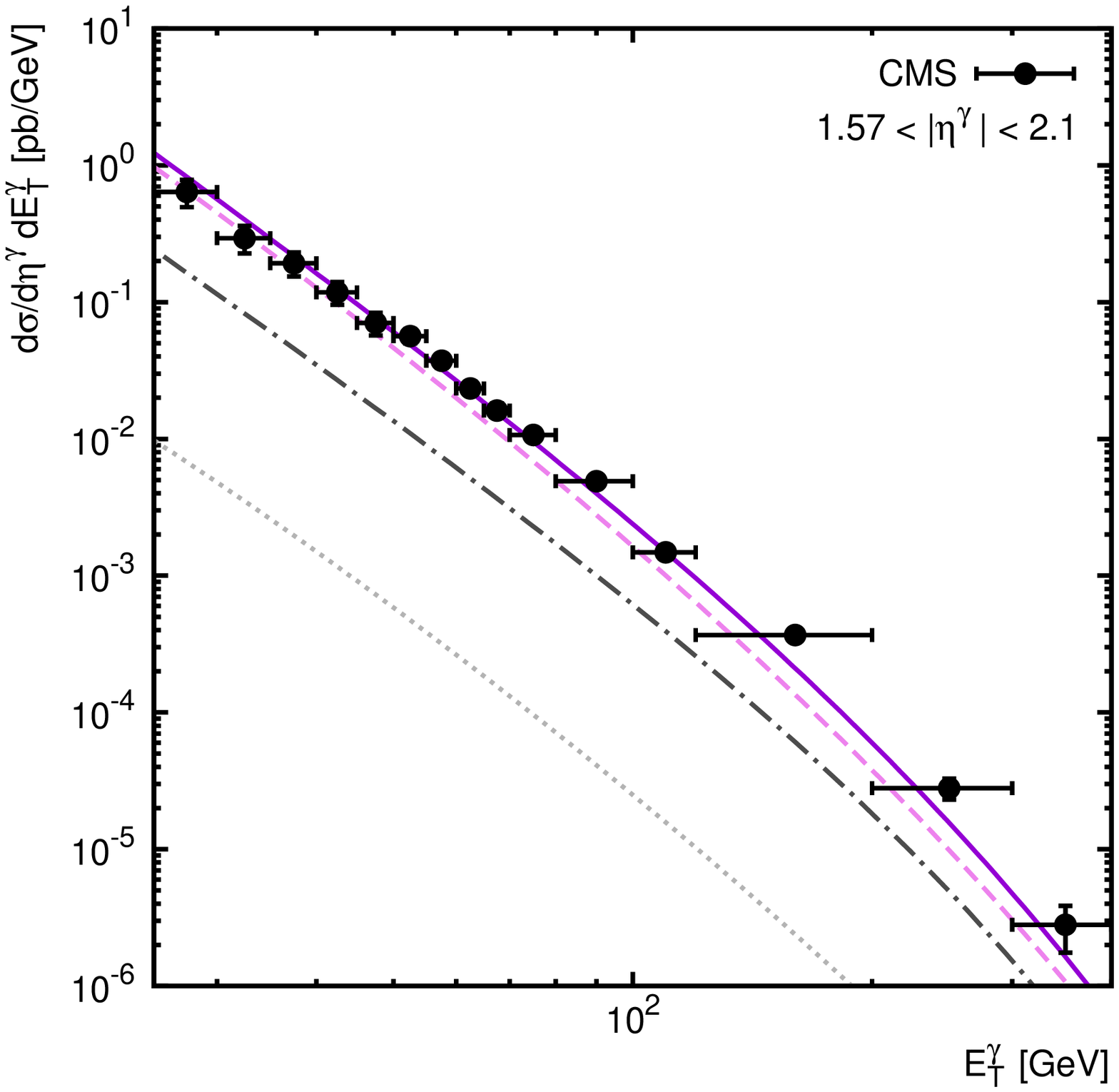, width = 6cm, angle = 270}
\hspace{-1cm}
\epsfig{figure=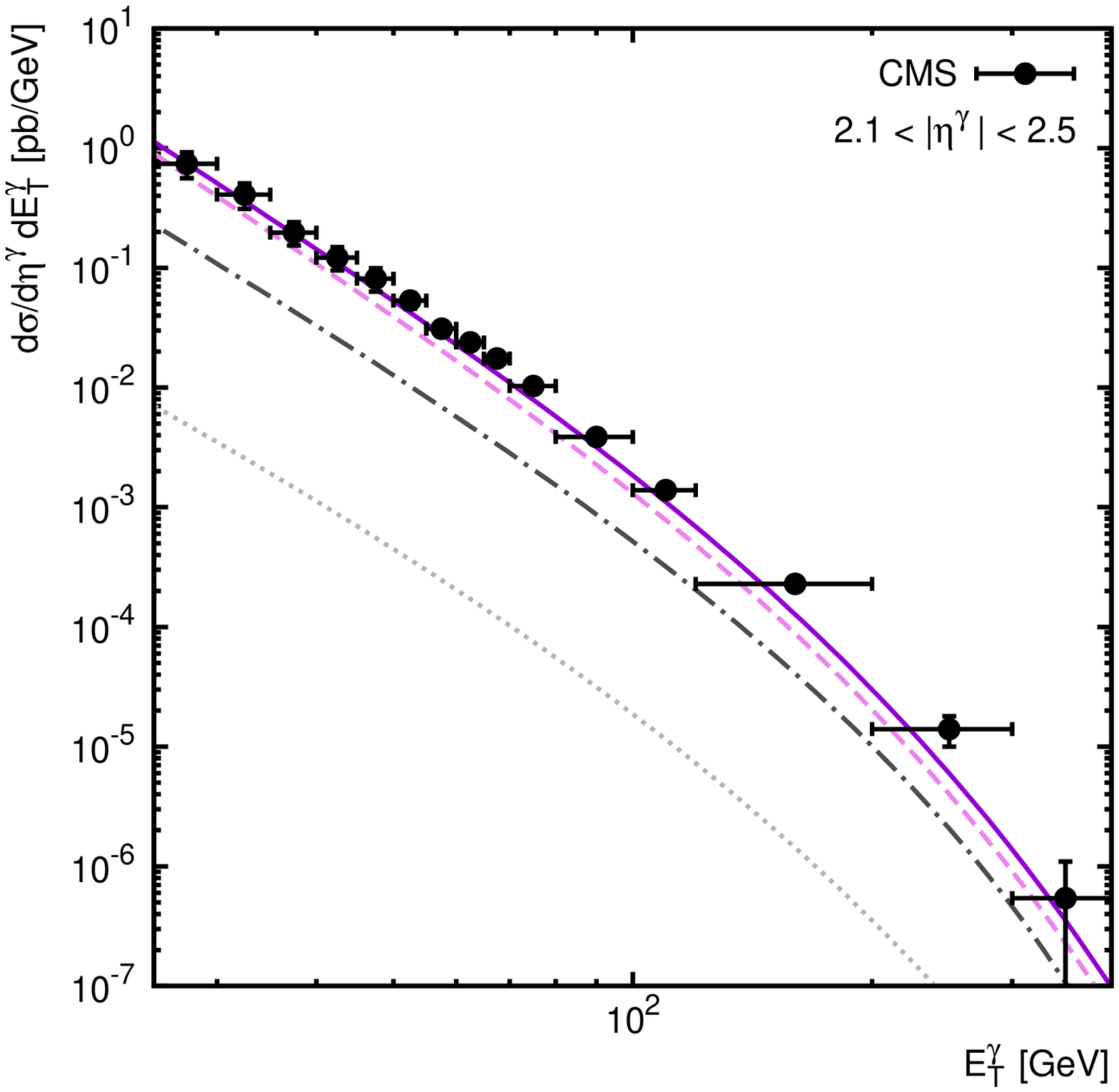, width = 6cm, angle = 270}
\caption{Different contributions to the inclusive prompt photon production 
cross sections at $\sqrt s = 7$~TeV. The dashed, dotted and dash-dotted 
curves correspond to the $q^* g^* \to \gamma q$, $q^* \bar q^* \to \gamma g$ and
$q q^\prime \to \gamma q q^\prime$ subprocesses.
The solid curves represent the sum of all contributions.
The experimental data are from CMS\cite{1}.}
\label{fig5}
\end{center}
\end{figure}

\begin{figure}
\begin{center}
\epsfig{figure=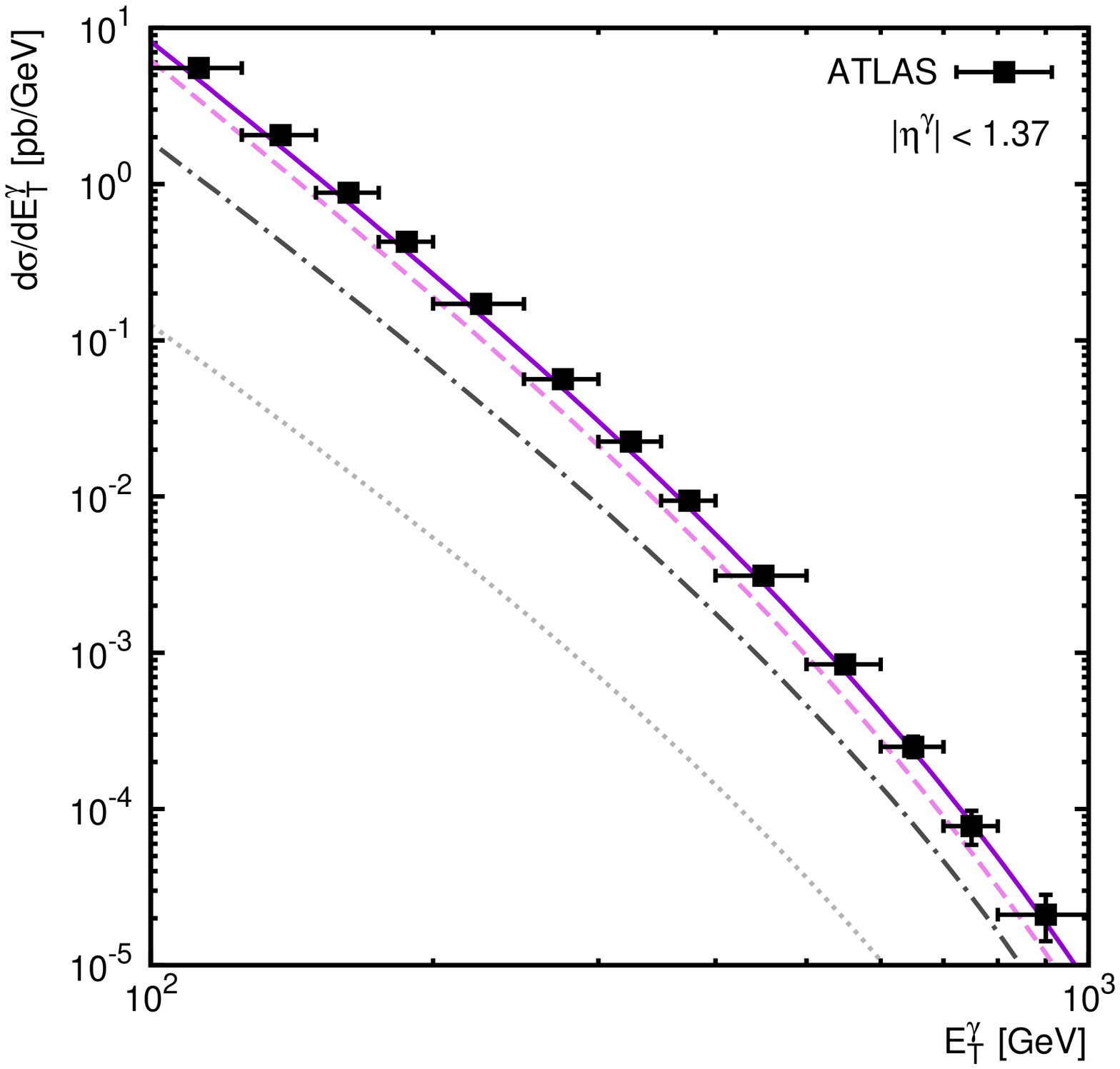, width = 5.9cm, angle = 270}
\hspace{-1cm}
\epsfig{figure=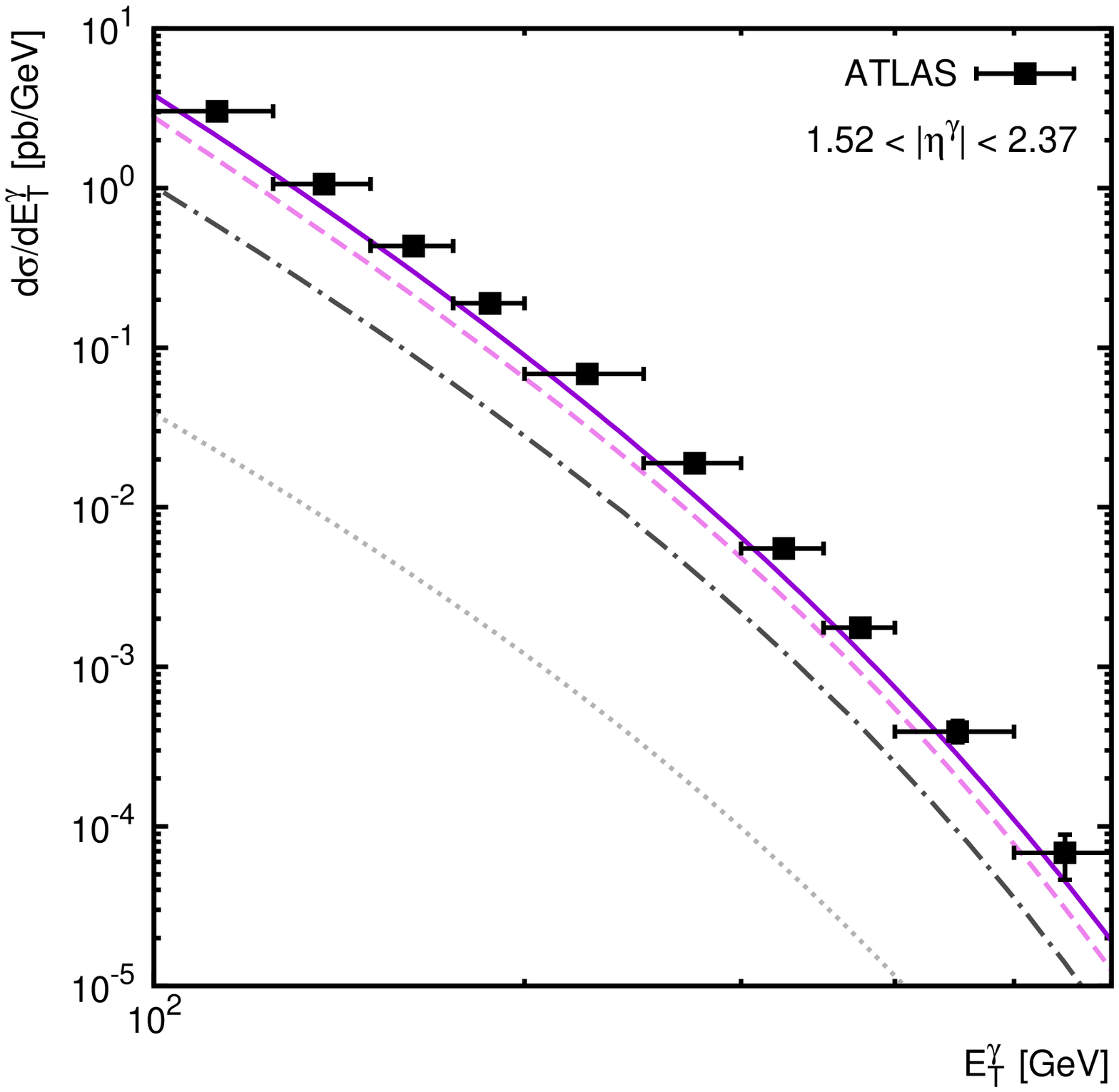, width = 5.9cm, angle = 270}
\caption{Different contributions to the inclusive prompt photon production 
cross sections at $\sqrt s = 7$~TeV. 
Notation of all curves is the same as in Fig.~5. 
The experimental data are from ATLAS\cite{2}.}
\label{fig6}
\end{center}
\end{figure}

\begin{figure}
\begin{center}
\epsfig{figure=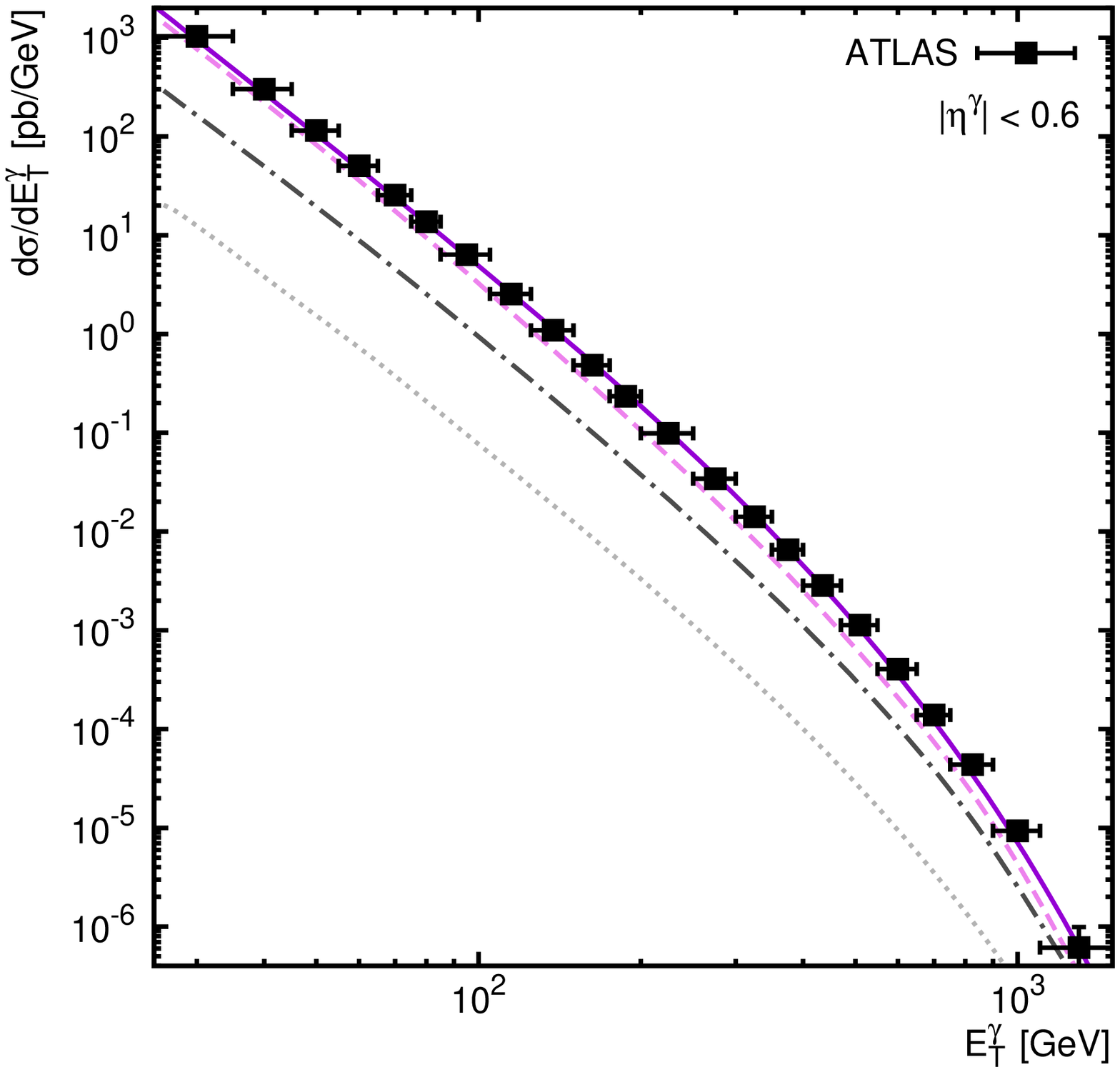, width = 6cm, angle = 270} 
\vspace{0.7cm} \hspace{-1cm}
\epsfig{figure=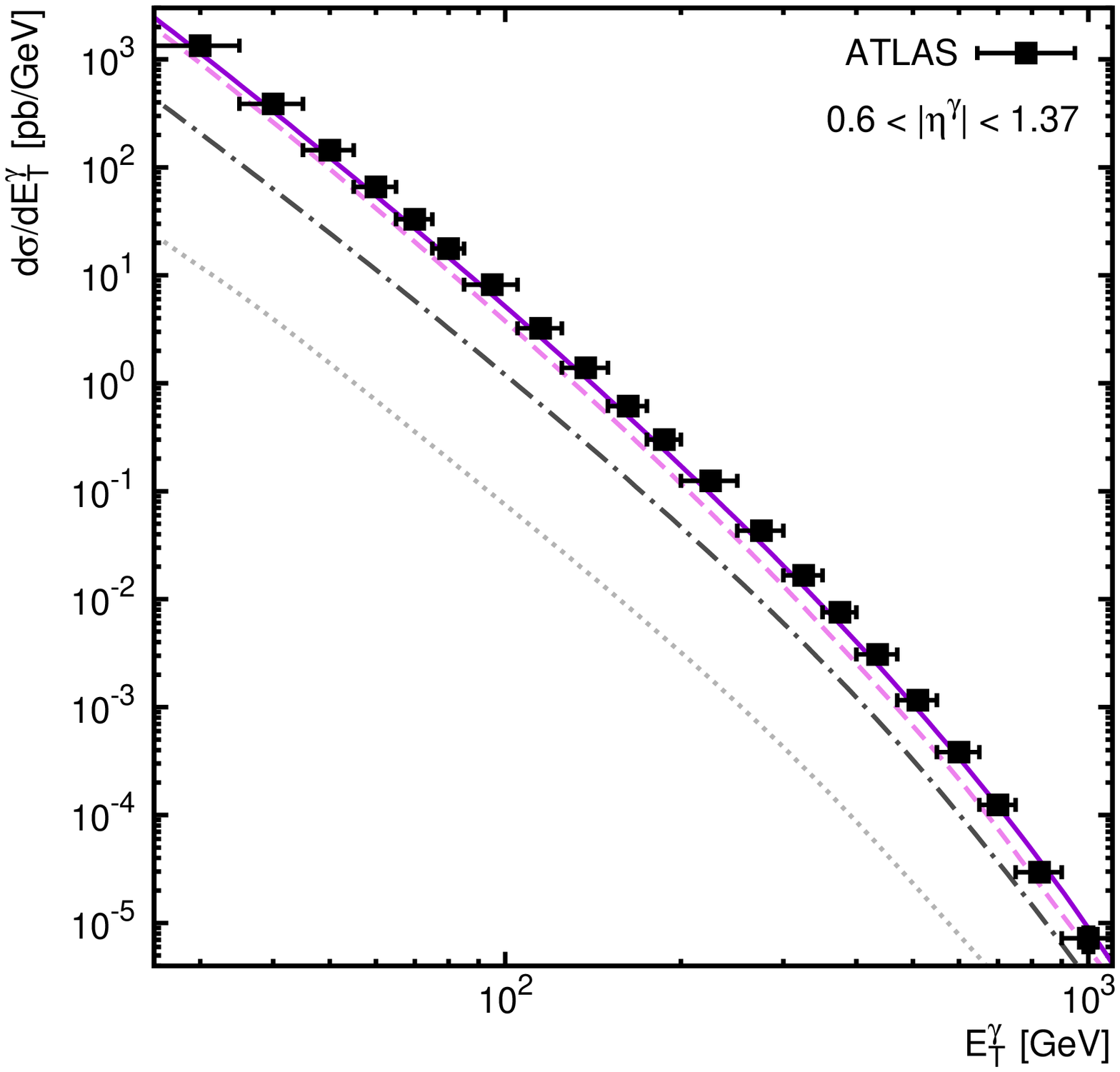, width = 6cm, angle = 270} 
\vspace{0.7cm}
\epsfig{figure=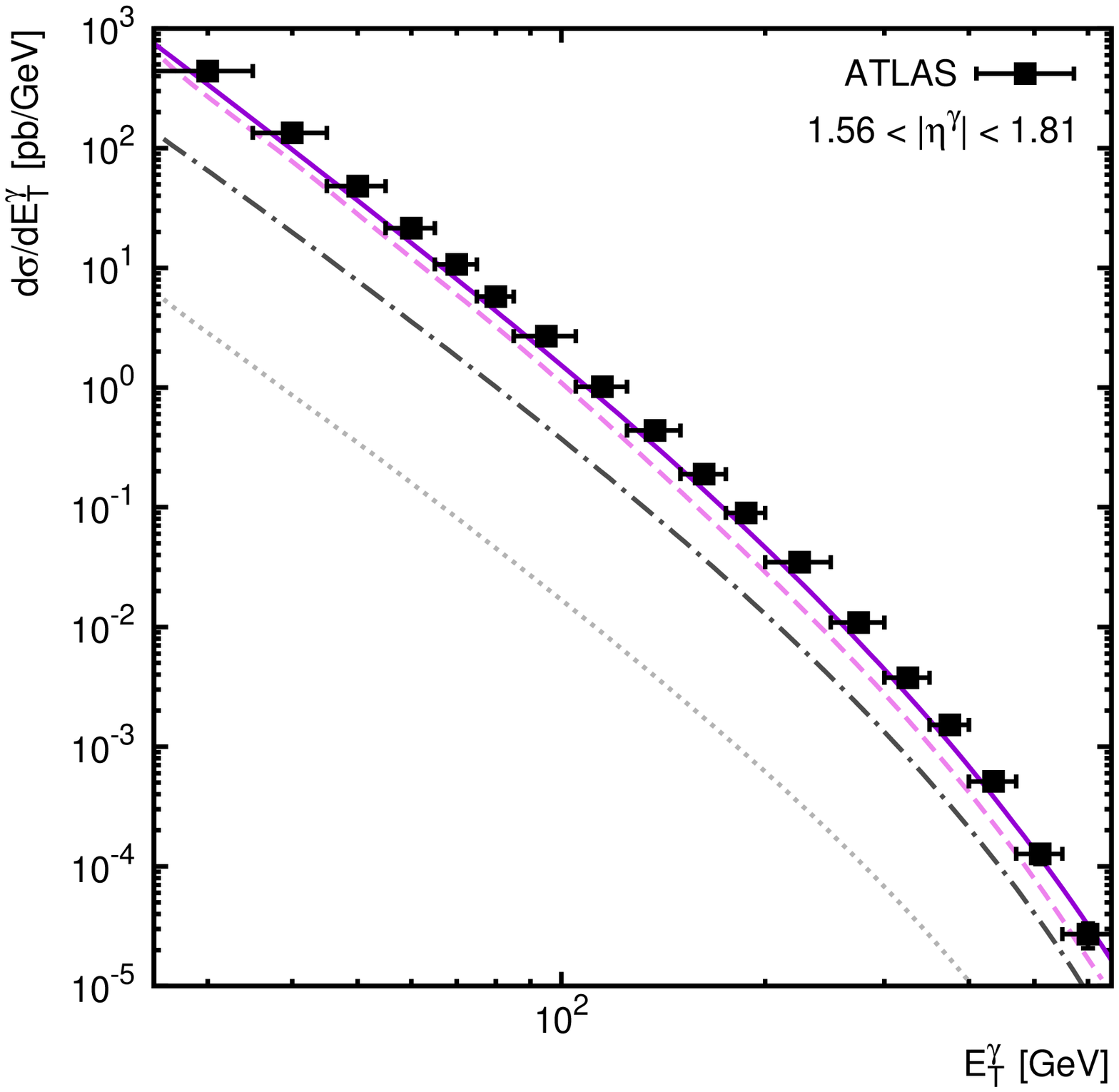, width = 6cm, angle = 270}
\hspace{-1cm}
\epsfig{figure=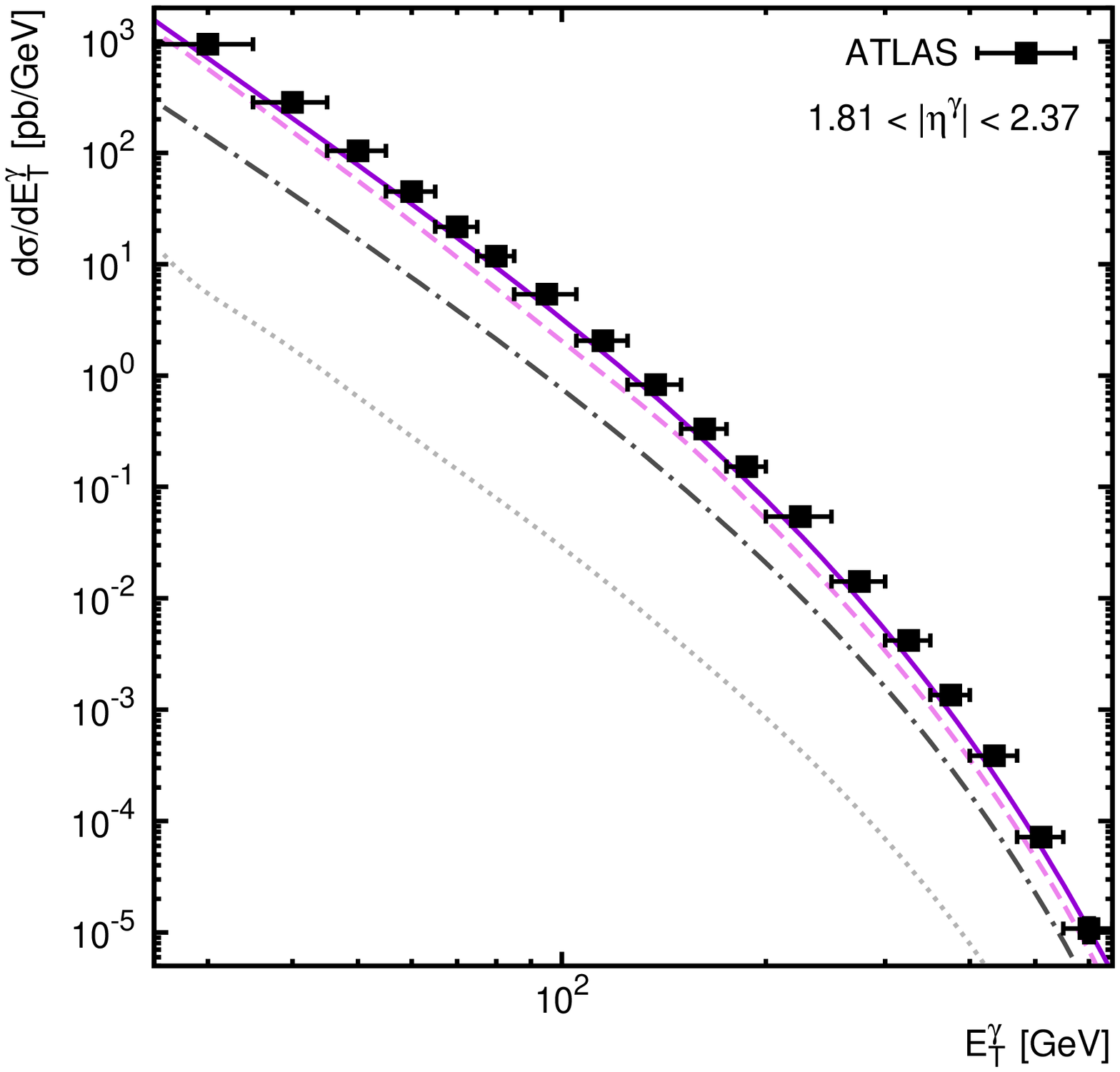, width = 6cm, angle = 270}
\caption{Different contributions to the inclusive prompt photon production 
cross sections at $\sqrt s = 8$~TeV. 
Notation of all curves is the same as in Fig.~5. 
The experimental data are from ATLAS\cite{3}.}
\label{fig7}
\end{center}
\end{figure}

\end{document}